# Commutator Matrix in Phase Space Mapping Models for Nonadiabatic Quantum Dynamics

*Xin He, Baihua Wu, Zhihao Gong, Jian Liu*[*]

Beijing National Laboratory for Molecular Sciences, Institute of Theoretical and Computational Chemistry, College of Chemistry and Molecular Engineering,

Peking University, Beijing 100871, China

[*] Author to whom correspondence should be addressed: jianliupku@pku.edu.cn

**KEYWORDS**: Nonadiabatic dynamics, Phase space mapping model, Meyer-Miller Hamiltonian, Trajectory-based quantum dynamics approach

**ABSTRACT**: We show that a novel, general phase space mapping Hamiltonian for nonadiabatic systems, which is reminiscent of the renowned Meyer-Miller mapping Hamiltonian, involves a commutator variable matrix rather than the conventional zero-point-energy parameter. In the exact mapping formulation on constraint space for phase space approaches for nonadiabatic dynamics, the general mapping Hamiltonian with commutator variables can be employed to generate approximate trajectory-based dynamics. Various benchmark model tests, which range from gas phase to condensed phase systems, suggest that the overall performance of the general mapping Hamiltonian is better than that of the conventional Meyer-Miller Hamiltonian.

1.  **Introduction**

Many important processes from photochemistry to electron transfer in chemical, biological, and materials systems involve quantum mechanical behavior of both electrons and nuclei in the context of nonadiabatic dynamics[1-10]. The celebrated Meyer-Miller mapping model[11-13] is one of the important theoretical frameworks for developing practical nonadiabatic dynamics methods[14-56]. Consider a coupled $F$-electronic-state Hamiltonian operator

$$\hat{H} = \sum_{n,m=1}^{F} H_{nm}(\hat{\mathbf{R}}, \hat{\mathbf{P}})|n\rangle\langle m| = \sum_{n,m=1}^{F}\left[\frac{1}{2}\hat{\mathbf{P}}^T\mathbf{M}^{-1}\hat{\mathbf{P}}\delta_{nm} + V_{nm}(\hat{\mathbf{R}})\right]|n\rangle\langle m| \quad (1)$$

in the diabatic representation (for simplicity), where the $F$ electronic states consist of an orthogonal complete basis set, i.e.,

$$\langle m|n\rangle = \delta_{mn}$$
$$\hat{I}_{ele} = \sum_{n=1}^{F}|n\rangle\langle n| \quad (2)$$

Here $\hat{I}_{ele}$ is the identity operator in the electronic state space, $\mathbf{M}$ is the diagonal 'mass matrix' with elements $\{m_j\}$, $\{\mathbf{R},\mathbf{P}\}$ are the coordinate and momentum variables for the nuclear DOFs (with $N$ the total number of nuclear DOFs), and potential energy elements $V_{nm}(\mathbf{R}) = V_{mn}(\mathbf{R})$ form a real symmetric matrix. The Meyer-Miller Hamiltonian reads

$$H_{\text{MM}}(\mathbf{R},\mathbf{P};\mathbf{x},\mathbf{p}) = \frac{1}{2}\mathbf{P}^T\mathbf{M}^{-1}\mathbf{P} + \sum_{n,m=1}^{F}\left[\frac{1}{2}(x^{(n)}x^{(m)} + p^{(n)}p^{(m)}) - \gamma\delta_{nm}\right]V_{nm}(\mathbf{R}) \quad , \quad (3)$$

where $\{\mathbf{x},\mathbf{p}\} = \{x^{(1)},\cdots,x^{(F)},p^{(1)},\cdots,p^{(F)}\}$ are the mapping coordinate and momentum variables for the $F$ electronic states.

There exist two alternative approaches that derive the Meyer-Miller mapping model in quantum mechanics[12, 13]. The approach of Stock and Thoss[12] and of Sun et. al[14] suggests that parameter $\gamma$ in the Meyer-Miller mapping Hamiltonian eq 3 is a parameter for the zero point energy of a singly excited oscillator for the underlying mapping DOFs for each electronic state[30-33], which is in the similar spirit to that of the pioneering work of Meyer and Miller[11]. In comparison, the unified framework proposed in Ref. [13] offers a substantially different picture. It derives a mapping model reminiscent of the Meyer-Miller model,

$$H_{\mathrm{MM}}\left(\hat{\mathbf{R}},\hat{\mathbf{P}};\mathbf{x},\mathbf{p}\right) = \sum_{n,m=1}^{F}\left[\frac{1}{2}(x^{(n)}x^{(m)} + p^{(n)}p^{(m)}) - \gamma\delta_{nm}\right]H_{mn}\left(\hat{\mathbf{R}},\hat{\mathbf{P}}\right) \quad . \quad (4)$$

where parameter $\gamma$ is interpreted as a parameter originated from a commutator of Pauli matrix $-\frac{i}{4}\left[\hat{\sigma}_x^{(n)},\hat{\sigma}_y^{(n)}\right] = -\frac{\hat{\sigma}_z^{(n)}}{2}$, which allows both positive and negative values[13, 42]. The one-to-one correspondence mapping formulation for the correlation function for nonadiabatic systems can rigorously be established[41, 42].

More interestingly, where the Meyer-Miller mapping Hamiltonian is re-derived in the novel framework in Ref. [13], it is also indicated that there exists a more general mapping Hamiltonian

$$H_{map}\left(\mathbf{R},\mathbf{P};\mathbf{x},\mathbf{p};\mathbf{\Gamma}\right) = \sum_{n,m=1}^{F}\left[\frac{1}{2}(x^{(n)}x^{(m)} + p^{(n)}p^{(m)}) - \Gamma_{nm}\right]H_{mn}\left(\mathbf{R},\mathbf{P}\right) \quad , \quad (5)$$

where $\mathbf{\Gamma}$ is a real symmetric matrix with the element $\Gamma_{nm}$ in the $n$-th row and $m$-th column. Its element $\Gamma_{nm}$ is a real variable for the corresponding commutator, $-\frac{i}{8}\left(\left[\hat{\sigma}_x^{(n)},\hat{\sigma}_y^{(m)}\right] + \left[\hat{\sigma}_x^{(m)},\hat{\sigma}_y^{(n)}\right]\right)$, as presented in Ref. [13]. $\mathbf{\Gamma}$ is denoted the commutator matrix. When off-diagonal elements of commutator matrix $\mathbf{\Gamma}$ are set to zero and diagonal elements are the same, the general mapping Hamiltonian eq 5 is simplified to eq 4. Note that commutator matrix $\mathbf{\Gamma}$ can evolve with time when eq 5 is utilized to generate corresponding Hamilton's equations of motion. To the best of our knowledge, except presented in Ref. [13], the general mapping Hamiltonian eq 5 with commutator matrix $\mathbf{\Gamma}$ as variables has *never* been proposed and used to generate the equations of motion for nonadiabatic dynamics. The purpose of the paper is to employ the general mapping Hamiltonian eq 5 for trajectory-based dynamics for nonadiabatic systems, in the exact mapping kernel formulation that we established in Refs. [41, 42]. The paper is organized as follows. Section 2 first reviews the one-to-one correspondence mapping formulation derived in Refs. [41, 42], then derives Hamilton's equations of motion from eq 5, where the frozen-nuclei limit as well as the Born-Oppenheimer limit are satisfied. Section 3 presents numerical results of various benchmark model tests for gas phase as well as condensed phase systems, which include the scattering models of Tully[57], 3-state photodissociation models of Miller and coworkers[58], 7-site model of the Fenna-Matthews-Olson (FMO) monomer[59], and atom-in-cavity models[60-63]. Finally, conclusions are given in Section 4.

## 2. Theory

### 2.1 Phase space mapping formulations for nonadiabatic systems

Because it is convenient to obtain useful insight about the correspondence between quantum and classical concepts in phase space formulations of quantum mechanics[64-82], they have been widely used in many areas of physics and chemistry since Wigner's pioneering work[64]. More recently, we have proposed a unified framework for the one-to-one correspondence mapping in phase space formulations of quantum mechanics[42], which naturally includes and surpasses the classification scheme[79, 81, 82] for conventional approaches[64-78, 80] for quantum systems represented in the continuous coordinate space, and is able to treat quantum systems described in the finite-dimensional Hilbert space[13, 41]. Such a framework offers a useful tool for nonadiabatic systems where both continuous nuclear degrees of freedom (DOFs) and discretized electronic state DOFs are involved.

In the unified framework of phase space mapping models for the (coupled) multi-state Hamiltonian (eq 1) in Ref. [13], eq 4 is the mapping model reminiscent of the Meyer-Miller model. When the mapping variables for the electronic state DOFs satisfy

$$\sum_{n=1}^{F}\left[\frac{(x^{(n)})^2+(p^{(n)})^2}{2}-\gamma\right]=1 \quad , \tag{6}$$

the mapping Hamiltonian eq 4 is equal to the conventional Meyer-Miller Hamiltonian eq 3. This was first proposed in Ref. [41] for general $F$-state systems. The simplest way is to use the full constraint electronic space that eq 6 defines, i.e.,

$$\mathcal{S}(\mathbf{x},\mathbf{p}) : \delta\left(\sum_{n=1}^{F}\left[\frac{(x^{(n)})^2+(p^{(n)})^2}{2}\right]-(1+F\gamma)\right) \quad , \tag{7}$$

for constructing the formulation for physical properties in the mapping approach. The possible value of parameter $\gamma$ for eq 6 or eq 7 implies $\gamma \in \left(-\frac{1}{F},\infty\right)$.

The trace of a product of two operators is expressed in phase space as

$$\text{Tr}_{n,e}[\hat{A}\hat{B}]=\int(2\pi\hbar)^{-N}d\mathbf{R}d\mathbf{P}\int_{\mathcal{S}(\mathbf{x},\mathbf{p})}Fd\mathbf{x}d\mathbf{p}A(\mathbf{R},\mathbf{P};\mathbf{x},\mathbf{p})\tilde{B}(\mathbf{R},\mathbf{P};\mathbf{x},\mathbf{p}) \tag{8}$$

where

$$A(\mathbf{R},\mathbf{P};\mathbf{x},\mathbf{p})=\text{Tr}_{n,e}\left[\hat{A}\hat{K}_{nuc}(\mathbf{R},\mathbf{P})\otimes\hat{K}_{ele}(\mathbf{x},\mathbf{p})\right] \quad , \tag{9}$$

$$\tilde{B}(\mathbf{R},\mathbf{P};\mathbf{x},\mathbf{p})=\text{Tr}_{n,e}\left[\hat{K}_{nuc}^{-1}(\mathbf{R},\mathbf{P})\otimes\hat{K}_{ele}^{-1}(\mathbf{x},\mathbf{p})\hat{B}\right] \quad , \tag{10}$$

$(2\pi\hbar)^{-N}d\mathbf{R}d\mathbf{P}\otimes Fd\mathbf{x}d\mathbf{p}$ stands for the invariant measure on the mapping phase space for nuclear and electronic state DOFs, and $\text{Tr}_n$ and $\text{Tr}_e$ represent the trace over the nuclear DOFs and that over the $F$ electronic states, respectively. The mapping kernel and its inverse satisfy the normalization

$$\mathrm{Tr}_n\left[\hat{K}_{nuc}(\mathbf{R},\mathbf{P})\right] = \mathrm{Tr}_n\left[\hat{K}_{nuc}^{-1}(\mathbf{R},\mathbf{P})\right] = 1$$
$$\mathrm{Tr}_e\left[\hat{K}_{ele}(\mathbf{x},\mathbf{p})\right] = \mathrm{Tr}_e\left[\hat{K}_{ele}^{-1}(\mathbf{x},\mathbf{p})\right] = 1 \tag{11}$$

and

$$(2\pi\hbar)^{-N}\int d\mathbf{R}d\mathbf{P}\,\hat{K}_{nuc}(\mathbf{R},\mathbf{P}) = \hat{I}_{nuc}$$
$$(2\pi\hbar)^{-N}\int d\mathbf{R}d\mathbf{P}\,\hat{K}_{nuc}^{-1}(\mathbf{R},\mathbf{P}) = \hat{I}_{nuc}$$
$$\int_{\mathcal{S}(\mathbf{x},\mathbf{p})} F d\mathbf{x}d\mathbf{p}\,\hat{K}_{ele}(\mathbf{x},\mathbf{p}) = \hat{I}_{ele}$$
$$\int_{\mathcal{S}(\mathbf{x},\mathbf{p})} F d\mathbf{x}d\mathbf{p}\,\hat{K}_{ele}^{-1}(\mathbf{x},\mathbf{p}) = \hat{I}_{ele} \tag{12}$$

where $\hat{I}_{nuc}$ is the identity operator in the nuclear space and the integral over constraint mapping space $\mathcal{S}(\mathbf{x},\mathbf{p})$ is

$$\int_{\mathcal{S}(\mathbf{x},\mathbf{p})} F d\mathbf{x}d\mathbf{p}\, g(\mathbf{x},\mathbf{p}) = \frac{\int F d\mathbf{x}d\mathbf{p}\,\delta\left(\sum_{n=1}^{F}\left[\frac{(x^{(n)})^2 + (p^{(n)})^2}{2}\right] - (1+F\gamma)\right) g(\mathbf{x},\mathbf{p})}{\int d\mathbf{x}d\mathbf{p}\,\delta\left(\sum_{n=1}^{F}\left[\frac{(x^{(n)})^2 + (p^{(n)})^2}{2}\right] - (1+F\gamma)\right)} . \tag{13}$$

The one-to-one correspondence mapping from $A(\mathbf{R},\mathbf{P};\mathbf{x},\mathbf{p})$ (or $\tilde{B}(\mathbf{R},\mathbf{P};\mathbf{x},\mathbf{p})$) of eq 9 back to operator $\hat{A}$ (or $\hat{B}$) is

$$\hat{A} = \int (2\pi\hbar)^{-N} d\mathbf{R}d\mathbf{P}\int_{\mathcal{S}(\mathbf{x},\mathbf{p})} F d\mathbf{x}d\mathbf{p}\, A(\mathbf{R},\mathbf{P};\mathbf{x},\mathbf{p})\hat{K}_{nuc}^{-1}(\mathbf{R},\mathbf{P})\otimes \hat{K}_{ele}^{-1}(\mathbf{x},\mathbf{p})$$
$$\hat{B} = \int (2\pi\hbar)^{-N} d\mathbf{R}d\mathbf{P}\int_{\mathcal{S}(\mathbf{x},\mathbf{p})} F d\mathbf{x}d\mathbf{p}\, \tilde{B}(\mathbf{R},\mathbf{P};\mathbf{x},\mathbf{p})\hat{K}_{nuc}(\mathbf{R},\mathbf{P})\otimes \hat{K}_{ele}(\mathbf{x},\mathbf{p}) \tag{14}$$

The mapping kernel for the nuclear DOFs (in eq 9) is

$$\hat{K}_{nuc}(\mathbf{R},\mathbf{P}) = \left(\frac{\hbar}{2\pi}\right)^N \int d\boldsymbol{\zeta}\int d\boldsymbol{\eta}\, e^{i\boldsymbol{\zeta}\cdot(\hat{\mathbf{R}}-\mathbf{R}) + i\boldsymbol{\eta}\cdot(\hat{\mathbf{P}}-\mathbf{P})} f(\boldsymbol{\zeta},\boldsymbol{\eta}) , \tag{15}$$

and its inverse is

$$\hat{K}_{nuc}^{-1}(\mathbf{R},\mathbf{P}) = \left(\frac{\hbar}{2\pi}\right)^N \int d\boldsymbol{\zeta}\int d\boldsymbol{\eta}\, e^{-i\boldsymbol{\zeta}\cdot(\hat{\mathbf{R}}-\mathbf{R}) - i\boldsymbol{\eta}\cdot(\hat{\mathbf{P}}-\mathbf{P})} \left[f(-\boldsymbol{\zeta},-\boldsymbol{\eta})\right]^{-1} , \tag{16}$$

where $f(\boldsymbol{\zeta},\boldsymbol{\eta})$ is a scalar function to determine the corresponding nuclear phase space. For instance, the Wigner function[64, 65] has

$$f(\boldsymbol{\zeta},\boldsymbol{\eta})=1 \quad, \tag{17}$$

and the Husimi function[68] has

$$f(\boldsymbol{\zeta},\boldsymbol{\eta})=\exp\left(-\frac{\boldsymbol{\zeta}^T \mathbf{G}^{-1}\boldsymbol{\zeta}}{4}-\frac{\hbar^2}{4}\boldsymbol{\eta}^T\mathbf{G}\boldsymbol{\eta}\right) \quad. \tag{18}$$

The mapping kernel for the $F$ electronic states (in eq 9) is

$$\hat{K}_{ele}(\mathbf{x},\mathbf{p})=\sum_{n,m=1}^{F}\left[\frac{1}{2}\left(x^{(n)}+ip^{(n)}\right)\left(x^{(m)}-ip^{(m)}\right)-\gamma\delta_{nm}\right]|n\rangle\langle m| \tag{19}$$

and the inverse kernel

$$\hat{K}_{ele}^{-1}(\mathbf{x},\mathbf{p})=\sum_{n,m=1}^{F}\left[\frac{1+F}{2(1+F\gamma)^2}\left(x^{(n)}+ip^{(n)}\right)\left(x^{(m)}-ip^{(m)}\right)-\frac{1-\gamma}{1+F\gamma}\delta_{nm}\right]|n\rangle\langle m| \quad. \tag{20}$$

In eqs 8, 12, 14, 15 and 16, while the integrals for the nuclear DOFs are over the whole nuclear phase space when the Wigner or Husimi function is employed, those for the $F$ electronic states are over the constraint electronic mapping space, $\mathcal{S}(\mathbf{x},\mathbf{p})$.

When the nuclear DOFs are described in Wigner phase space (eqs 15-16 with eq 17), the mapping kernel and its inverse are the same, i.e.,

$$\hat{K}_{nuc}(\mathbf{R},\mathbf{P})=\hat{K}_{nuc}^{-1}(\mathbf{R},\mathbf{P}) \quad. \tag{21}$$

When the Wigner function (eqs 15-16 with eq 17) is used for the nuclear DOFs, it is easy to show that the real part of the mapping Hamiltonian $H(\mathbf{R},\mathbf{P};\mathbf{x},\mathbf{p})=\mathrm{Tr}_{n,e}\left[\hat{H}\hat{K}_{nuc}(\mathbf{R},\mathbf{P})\otimes\hat{K}_{ele}(\mathbf{x},\mathbf{p})\right]$ with constraint eq 7 for the coupled multi-state Hamiltonian operator eq 1 is the same as the conventional Meyer-Miller Hamiltonian eq 3. The equations of motion produced by the real part of $H(\mathbf{R},\mathbf{P};\mathbf{x},\mathbf{p})=\text{Tr}_{n,e}\left[\hat{H}\hat{K}_{nuc}(\mathbf{R},\mathbf{P})\otimes\hat{K}_{ele}(\mathbf{x},\mathbf{p})\right]$ are identical to those yielded by its imaginary part.

When parameter

$$\gamma = \frac{\sqrt{F+1}-1}{F} \qquad (22)$$

is employed, the mapping kernel for the electronic DOFs is equal to the inverse kernel, i.e.,

$$\hat{K}_{ele}(\mathbf{x},\mathbf{p}) = \hat{K}_{ele}^{-1}(\mathbf{x},\mathbf{p}) \quad . \qquad (23)$$

Equation 22 offers the only physical value for parameter $\gamma$ in region $\left(-\frac{1}{F},\infty\right)$ to make eq 23 hold. We note that the so called spin mapping model of Refs. [43, 44] intrinsically based on the Meyer-Miller mapping Hamiltonian model (especially when $F \geq 3$ electronic states are involved) is only a special case of the exact phase space mapping formulation that we established first in Refs. [13] and [41] and then in Ref. [42], i.e., parameter $\gamma = 0, \left(\sqrt{F+1}-1\right)/F$, or 1 in our exact phase space mapping formulation corresponds to the Q-version, W-version, or P-version of Refs. [43, 44], respectively. Interestingly, the authors of Ref. [44] even failed to understand that the interpretation for general $F$-state systems constructed in Appendix A of Ref. [41] is simply an exact phase space mapping formulation for parameter $\gamma = 0$. We also note that the exact phase space mapping formulation of the correlation function can be used to formalize various other methods based on the Meyer-Miller mapping Hamiltonian model in Refs. [14, 35, 45, 46, 49, 83-86], as we will show in a forthcoming paper.

## 2.2 Expression of the time correlation function

Define the Heisenberg operator $\hat{B}(t) = e^{i\hat{H}t/\hbar}\hat{B}e^{-i\hat{H}t/\hbar}$. As a result of eq 8, an exact expression of the time correlation function of the nonadiabatic system

$$C_{AB}(t) = \text{Tr}_{n,e}\left[\hat{A}(0)\hat{B}(t)\right] \tag{24}$$

is

$$C_{AB}(t) = \int (2\pi\hbar)^{-N} d\mathbf{R} d\mathbf{P} \int_{\mathcal{S}(\mathbf{x},\mathbf{p})} F d\mathbf{x} d\mathbf{p} A(\mathbf{R},\mathbf{P};\mathbf{x},\mathbf{p}) \tilde{B}(\mathbf{R},\mathbf{P};\mathbf{x},\mathbf{p};t) , \tag{25}$$

where

$$\tilde{B}(\mathbf{R},\mathbf{P};\mathbf{x},\mathbf{p};t) = \text{Tr}_{n,e}\left[\hat{K}_{nuc}^{-1}(\mathbf{R},\mathbf{P}) \otimes \hat{K}_{ele}^{-1}(\mathbf{x},\mathbf{p})\hat{B}(t)\right] . \tag{26}$$

When nuclear and electronic dynamics is exactly solved in eq 26, the correlation function formulation eq 25 is exact for nonadiabatic systems[41, 42].

When trajectory-based dynamics is introduced, eq 25 is recast into

$$C_{AB}(t) = \int (2\pi\hbar)^{-N} d\mathbf{R} d\mathbf{P} \int_{\mathcal{S}(\mathbf{x},\mathbf{p})} F d\mathbf{x} d\mathbf{p} A(\mathbf{R},\mathbf{P};\mathbf{x},\mathbf{p}) \tilde{B}(\mathbf{R}_t,\mathbf{P}_t;\mathbf{x}_t,\mathbf{p}_t) . \tag{27}$$

In the frozen-nuclei limit where only the electronic DOFs are involved (i.e., nuclear coordinate $\mathbf{R}$ and nuclear momentum $\mathbf{P}$ are fixed), Hamilton's equations of motion from the Meyer-Miller Hamiltonian eq 3 lead to exact results. When both nuclear and electronic DOFs are considered, exact equations of motion for the trajectories are far from trivial to solve numerically. When the independent trajectory is introduced to eq 27, it is then often an approximation to eq 25. For instance, when the Meyer-Miller Hamiltonian eq 3 is used to generate the independent trajectory for both nuclear and electronic DOFs in eq 27, it is equivalent to the extended classical mapping model (eCMM) approach[41, 42] where the linearized semiclassical or linearized path integral approximation[14, 87, 88] is utilized for only the nuclear DOFs.

Note that the one-to-one correspondence mapping framework for eq 8 as well as eq 25 only depends on the constraint phase space defined by eq 7. That is, the exact mapping framework is intrinsically *independent* of the form of the mapping Hamiltonian for dynamics. The Meyer-Miller Hamiltonian eq 3 is not necessary to be the *only* choice for yielding the equations of motion for the independent trajectory for eq 27. The derivation procedure for eq 43 for Model II of Ref. [13] suggests that a more comprehensive form for the phase space mapping Hamiltonian model is eq 5. When the equality

$$\sum_{n=1}^{F}\left[\frac{\left(x^{(n)}\right)^2+\left(p^{(n)}\right)^2}{2}-\Gamma_{nn}\right]=1 \qquad (28)$$

holds, eq 5 becomes

$$H_{map}(\mathbf{R},\mathbf{P};\mathbf{x},\mathbf{p};\mathbf{\Gamma})=\frac{1}{2}\mathbf{P}^T\mathbf{M}^{-1}\mathbf{P}+\sum_{n,m=1}^{F}\left[\frac{1}{2}(x^{(n)}x^{(m)}+p^{(n)}p^{(m)})-\Gamma_{nm}\right]V_{mn}(\mathbf{R}) \ . \qquad (29)$$

Any Hermitian Matrix $\mathbf{\Gamma}$ can be represented by its eigenvalues $\{\lambda_k\}$ and eigenvectors $\{\mathbf{b}_k\}$, i.e.,

$$\mathbf{\Gamma}=\sum_{k=1}^{F}\lambda_k \mathbf{b}_k \mathbf{b}_k^\dagger \ , \qquad (30)$$

with $\mathbf{\Gamma}\mathbf{b}_k=\lambda_k\mathbf{b}_k$ and symbol † standing for the complex conjugate transpose. As its eigenvalues $\{\lambda_k\}$ are real, define $\mathbf{c}_k=|\lambda_k|^{1/2}\mathbf{b}_k$, and eq 30 can be recast into

$$\mathbf{\Gamma}=\sum_{k=1}^{F}s_k \mathbf{c}_k \mathbf{c}_k^\dagger \ . \qquad (31)$$

Here, parameter

$$s_k=\text{sgn}(\lambda_k) \qquad (32)$$

is the sign of the eigenvalue $\lambda_k$. Using

$$c_k^{(n)}=\left(\tilde{x}_k^{(n)}+i\tilde{p}_k^{(n)}\right)/\sqrt{2} \qquad (33)$$

to represent the $n$-th element of vector $\mathbf{c}_k$, we obtain

$$\Gamma_{nm}=\sum_{k=1}^{F}\frac{s_k}{2}\left(\tilde{x}_k^{(n)}+i\tilde{p}_k^{(n)}\right)\left(\tilde{x}_k^{(m)}-i\tilde{p}_k^{(m)}\right) \ . \qquad (34)$$

Here, $\tilde{x}_k^{(n)}$ and $\tilde{p}_k^{(n)}$ are two real auxiliary variables. Because commutator matrix of eq 29 is a real symmetric matrix, eq 34 is simplified to

$$\Gamma_{nm} = \sum_{k=1}^{F} \frac{s_k}{2} \left( \tilde{x}_k^{(n)} \tilde{x}_k^{(m)} + \tilde{p}_k^{(n)} \tilde{p}_k^{(m)} \right) . \qquad (35)$$

The mapping Hamiltonian eq 29 then becomes

$$\begin{aligned} H_{map}(\mathbf{R},\mathbf{P};\mathbf{x},\mathbf{p};\tilde{\mathbf{x}},\tilde{\mathbf{p}}) &= \frac{1}{2}\mathbf{P}^T \mathbf{M}^{-1} \mathbf{P} \\ &+ \sum_{n,m=1}^{F} V_{mn}(\mathbf{R}) \left[ \frac{1}{2}(x^{(n)}x^{(m)} + p^{(n)}p^{(m)}) - \sum_{k=1}^{F} \frac{s_k}{2}\left(\tilde{x}_k^{(n)}\tilde{x}_k^{(m)} + \tilde{p}_k^{(n)}\tilde{p}_k^{(m)}\right) \right] \end{aligned} . \qquad (36)$$

The equations of motion governed by eq 36 are

$$\begin{aligned} \dot{\mathbf{R}} &= \mathbf{M}^{-1}\mathbf{P} \\ \dot{\mathbf{P}} &= -\sum_{n,m=1}^{F} \frac{\partial V_{mn}(\mathbf{R})}{\partial \mathbf{R}} \left[ \frac{1}{2}(x^{(n)}x^{(m)} + p^{(n)}p^{(m)}) - \sum_{k=1}^{F} \frac{s_k}{2}\left(\tilde{x}_k^{(n)}\tilde{x}_k^{(m)} + \tilde{p}_k^{(n)}\tilde{p}_k^{(m)}\right) \right] \end{aligned} . \qquad (37)$$

for nuclear DOFs, and

$$\begin{aligned} \dot{x}^{(n)} &= \sum_{m=1}^{F} V_{nm}(\mathbf{R}) p^{(m)} \\ \dot{p}^{(n)} &= -\sum_{m=1}^{F} V_{nm}(\mathbf{R}) x^{(m)} \\ \dot{\tilde{x}}_k^{(n)} &= -s_k \sum_{m=1}^{F} V_{nm}(\mathbf{R}) \tilde{p}_k^{(m)} \\ \dot{\tilde{p}}_k^{(n)} &= s_k \sum_{m=1}^{F} V_{nm}(\mathbf{R}) \tilde{x}_k^{(m)} \end{aligned} . \qquad (38)$$

for the electronic mapping DOFs $\{\mathbf{x},\mathbf{p}\}$ and auxiliary variables $\{\tilde{\mathbf{x}},\tilde{\mathbf{p}}\}$ for the commutator matrix. We denote this scheme the extended classical mapping model with commutator variables (eCMMcv). Because eq 28 holds at the beginning, it is straightforward to verify that the equations of motion (eqs 37-38) generated from the mapping Hamiltonian eq 36 conserve the two properties,

$$\sum_{n=1}^{F}\left[\frac{\left(x^{(n)}\right)^2+\left(p^{(n)}\right)^2}{2}\right]=(1+F\gamma) \tag{39}$$

that is equivalent to eq 7, and

$$\sum_{n,k=1}^{F}\frac{s_k}{2}\left(\left(\tilde{x}_k^{(n)}\right)^2+\left(\tilde{p}_k^{(n)}\right)^2\right)=F\gamma \quad. \tag{40}$$

Because eq 39 holds, the exact mapping framework for eq 8 as well as eq 25 for the electronic DOFs is still valid.

Further consider that the initial electronic state is localized at state $|j_{occ}\rangle$. While the initial condition for $(\mathbf{x},\mathbf{p})$ is uniformly sampled on constraint space $\mathcal{S}(\mathbf{x},\mathbf{p})$ that depicts the mapping framework for the electronic DOFs, the initial condition for auxiliary variables $(\tilde{\mathbf{x}},\tilde{\mathbf{p}})$ for commutator matrix $\mathbf{\Gamma}$ is given by

$$\frac{s_k}{2}\left(\left(\tilde{x}_k^{(n)}(0)\right)^2+\left(\tilde{p}_k^{(n)}(0)\right)^2\right)=\left[\frac{\left(x^{(n)}(0)\right)^2+\left(p^{(n)}(0)\right)^2}{2}-\delta_{n,j_{occ}}\right]\delta_{nk} \tag{41}$$

with

$$s_k=\begin{cases}-1 & \text{when } \dfrac{\left(x^{(k)}(0)\right)^2+\left(p^{(k)}(0)\right)^2}{2}<1 \text{ and } k=j_{occ}\\ 1 & \text{elsewhere}\end{cases} \tag{42}$$

such that the equations of motion of eqs 37-38 approach the Born-Oppenheimer limit for each trajectory when state-state coupling terms $\{V_{nm}(\mathbf{R})\}$ vanish for all $n\neq m$, i.e., it yields

$$\left[\frac{1}{2}\left(\left(x^{(n)}\right)^2+\left(p^{(n)}\right)^2\right)-\sum_{k=1}^{F}\frac{s_k}{2}\left(\left(\tilde{x}_k^{(n)}\right)^2+\left(\tilde{p}_k^{(n)}\right)^2\right)\right]=\delta_{n,j_{occ}} \quad. \tag{43}$$

The strategy similar to eq 43 has already been used in a few approaches based on the Meyer-Miller mapping Hamiltonian[29, 89, 90]. Because eqs 41-42 satisfy eq 28, the comprehensive phase space mapping Hamiltonian proposed by eq 5 is identical to eq 29 as well as eq 36. Equation 41 defines the constraint space for the initial conditions for auxiliary variables $(\tilde{\mathbf{x}}, \tilde{\mathbf{p}})$ for commutator matrix $\mathbf{\Gamma}$,

$$\xi(\tilde{\mathbf{x}}(0), \tilde{\mathbf{p}}(0)) : \prod_{k,n=1}^{F} \delta\left(\frac{s_k}{2}\left(\left(\tilde{x}_k^{(n)}(0)\right)^2 + \left(\tilde{p}_k^{(n)}(0)\right)^2\right) - \left[\frac{\left(x^{(n)}(0)\right)^2 + \left(p^{(n)}(0)\right)^2}{2} - \delta_{n,j_{occ}}\right]\delta_{nk}\right)$$

(44)

When the initial values for $(\mathbf{R}, \mathbf{P}; \mathbf{x}, \mathbf{p})$ are the same, any point on constraint space $\xi(\tilde{\mathbf{x}}, \tilde{\mathbf{p}})$ as the initial values for auxiliary variables $(\tilde{\mathbf{x}}, \tilde{\mathbf{p}})$ for matrix $\mathbf{\Gamma}$ leads to the same values for $(\mathbf{R}_t, \mathbf{P}_t; \mathbf{x}_t, \mathbf{p}_t)$ at time $t$ along the trajectory yielded by the equations of motion eqs 37-38. Note that only the initial values for $(\mathbf{R}, \mathbf{P}; \mathbf{x}, \mathbf{p})$ and the values for $(\mathbf{R}_t, \mathbf{P}_t; \mathbf{x}_t, \mathbf{p}_t)$ at time $t$ are employed in the correlation function, eq 27. Once that the initial values for physical variables $(\mathbf{R}, \mathbf{P}; \mathbf{x}, \mathbf{p})$ are given, we only choose a specific point on constraint space $\xi(\tilde{\mathbf{x}}, \tilde{\mathbf{p}})$ as the initial values for auxiliary variables $(\tilde{\mathbf{x}}, \tilde{\mathbf{p}})$, because it is *not* necessary to do sampling on constraint space $\xi(\tilde{\mathbf{x}}, \tilde{\mathbf{p}})$. It is easy to prove that the equations of motion of eqs 37-38 also approach the frozen-nuclei limit when only the electronic DOFs are involved (i.e., nuclear coordinate $\mathbf{R}$ and nuclear momentum $\mathbf{P}$ are fixed). That is, eq 35 is equivalent to solving the time-dependent Schrödinger equation for the electronic DOFs when the nuclear DOFs are frozen. A more convenient and equivalent way to evolve electronic maping DOFs and auxiliary variables in eCMMcv is to treat them in the matrix representation, which is provided in **Section S1-A** of Supporting Information.

In summary, the expression of the correlation function eq 27 on constraint space $\mathcal{S}(\mathbf{x}, \mathbf{p})$ for the electronic DOFs and Wigner phase space for the nuclear DOFs, trajectory-based dynamics (eqs 37-38) governed by the general mapping Hamiltonian eq 36, and initial values for auxiliary variables $(\tilde{\mathbf{x}},\tilde{\mathbf{p}})$ defined by a point on constraint space $\xi(\tilde{\mathbf{x}},\tilde{\mathbf{p}})$ are all necessary elements for eCMMcv, the phase space mapping approach for nonadiabatic dynamics which we propose in this paper. The eCMMcv approach meets the frozen-nuclei limit as well as the Born-Oppenheimer limit.

Finally, it is straightforward to express eCMMcv (or eCMM) in the adiabatic representation or other representations. When the general mapping Hamiltonian with the commutator matrix is used in eCMMcv to yield the equations of motion, they are similar to the equations of motion generated by the Meyer-Miller Hamiltonian. The strategy of Cotton and Miller in Ref. [26] can directly be extended to the general mapping Hamiltonian in the adiabatic representation. (See Section S1 of Supporting Information for more discussion.)

## 3. Results and discussions

Below we test the numerical performance of eCMMcv for a few illustrative benchmark gas phase and condensed phase systems. We first apply eCMMcv to Tully's scattering models[57] that contain single avoided crossing (SAC) and dual avoided crossing (DAC) examples. The second application consists of photo-dissociation models of Miller and coworkers[58], where more-realistic Morse potentials are involved. We then test typical system-bath models for condensed phase dissipative systems[91], which include the 7-state Fenna-Matthews-Olson (FMO) monomer that appears in photosynthesis in green sulfur bacteria[59, 92], and strongly coupled optical cavity-molecular matter systems used to control and manipulate chemical and physical processes[50, 60-63, 93].

### 3.1. Tully's scattering models

Tully's scattering models[57] are often used as benchmark applications to test nonadiabatic dynamics methods. The SAC and DAC models that mimic the surface intersection in molecular systems have widely been tested for mapping model dynamics[18, 21, 49, 94].

Tully's scattering problems are described by a two-state Hamiltonian (with the form of eq 1) with an atom of mass $m = 2000$ a. u. After scattering in the interaction region, the system evolves in plateau regions where diabatic potential functions $V_{nn}(R \to \infty)$ and $V_{nn}(R \to -\infty)$ are flat. The transmission and reflection coefficients are calculated for state $n$. In each eCMM or eCMMcv simulation, fully converged results are obtained by an ensemble average over 96,000 trajectories.

**3.1.1 Single avoided crossing**

In the SAC model, the diagonal elements of the potential operator are $V_{11} = -V_{22} = A(1-e^{-B|R|})\text{sgn}(R)$ and off-diagonal ones are $V_{12} = V_{21} = Ce^{-DR^2}$. The parameters (using atomic units) are $A = 0.01$, $B = 1.6$, $C = 0.005$, and $D = 1.0$. The initial condition follows an occupation on the state $1$ with the initial nuclear wavepacket $\Psi(R;t=0) \propto \exp[-\alpha(R-R_0)^2/2 + i(R-R_0)P_0/\hbar]$, where the Gaussian width parameter is $\alpha = 1$ a. u., the initial average coordinate is $R_0 = -3.8$ a. u., and the initial average momentum is $P_0$. The initial Wigner distribution for the nuclear coordinate is then $\rho_W^{\text{nuc}}(R,P) \propto \exp[-\alpha(R-R_0)^2 - (P-P_0)^2/(\alpha\hbar^2)]$.

In the simulations the initial average momentum $P_0$ ranges from 2 a. u. to 50 a. u.. We present numerical results for the transmission coefficients in Figure 1 and Figure 2. While Panels 1(a) and 1(b) demonstrate the transmission coefficient of state 1 and that of state 2, respectively, for eCMM in the diabatic representation, Panels 1(c) and 1(d) show those calculated by eCMM in the adiabatic representation. Figure 2 then presents those results obtained by eCMMcv. Two values $\gamma = \left(\sqrt{F+1}-1\right)/F = 0.366$ and $\gamma = 1/2$ are used for parameter $\gamma$ in eq 7 for eCMM and that in eq 39 for eCMMcv. Figures 1 and 2 demonstrate that either eCMM or eCMMcv yields results close to the exact transmission of either of state 1 and state 2 in a large range of the initial average momentum ( $P_0 \geq 10$ a. u. ), independent of the choice of $\gamma$ and of the representation of the electronic states. In the small initial average momentum region ( $P_0 < 10$ a. u. ), the exact results produced by the discrete variable representation (DVR) approach[95] exhibit a threshold. Both eCMM and eCMMcv are competent in capturing such a threshold. Comparison between eCMM (Figure 1) and eCMMcv (Figure 2) shows that eCMMcv is less sensitive to the value of parameter $\gamma$. (More comparison is available in **Section S2** of Supporting Information.) The regime $\left[\left(\sqrt{F+1}-1\right)/F, 1/2\right]$ is recommended for $\gamma$ in eCMMcv. We then focus on this reasonable regime for parameter $\gamma$, which will be tested for the rest of the benchmark models in the paper.

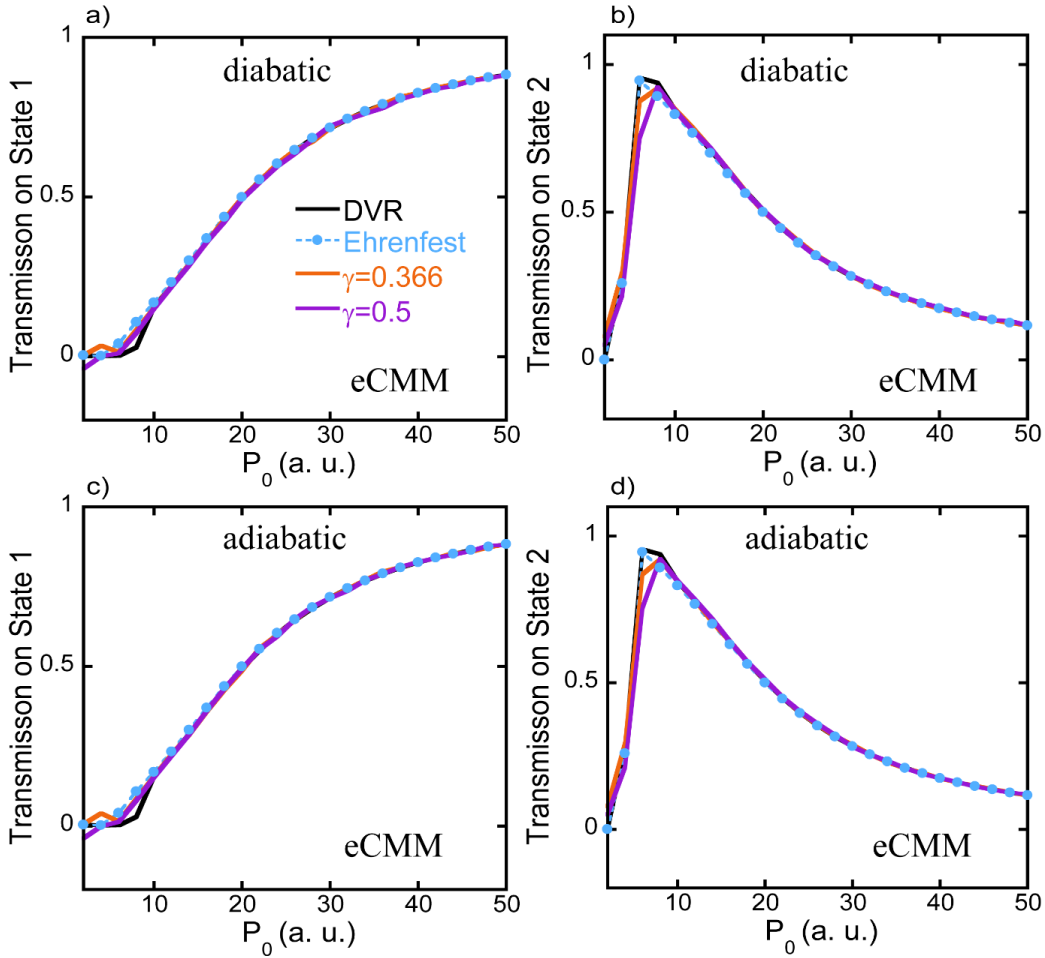

**Figure 1:** Transmission coefficients of Tully's SAC model. In panel a, Orange and purple solid lines: The eCMM transmission coefficients on state 1 for $\gamma = 0.366$ and $\gamma = 0.5$, respectively. Black solid line: Exact results yielded by DVR. Blue dashed line with circles: Ehrenfest dynamics. Panel b is similar to Panel a but for the eCMM transmission coefficients on state 2. Panels c and d are the same as panels a and b, respectively, except that the adiabatic representation is used in panels c and d. The transmission coefficients in the adiabatic representation are transformed to those in the diabatic representation for fair comparison.

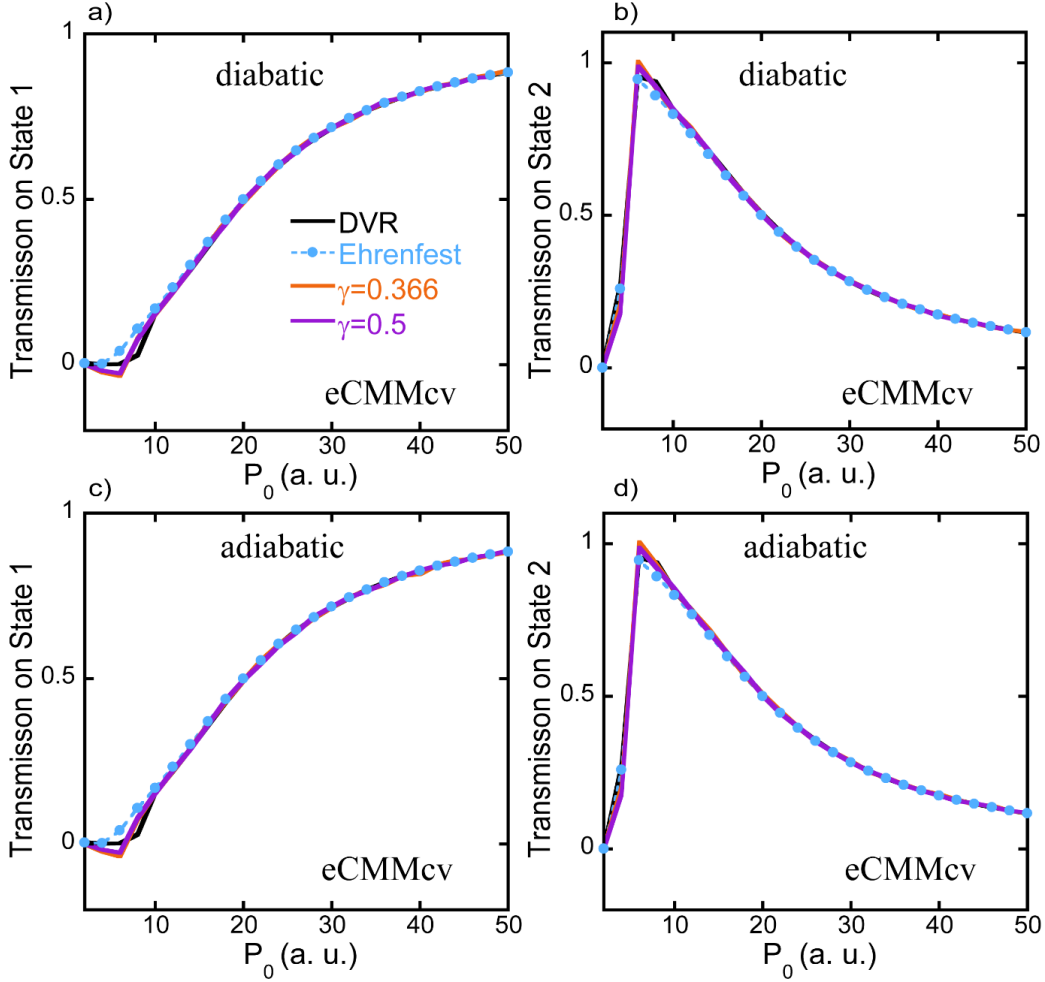

**Figure 2:** Same as Figure 1, but for transmission coefficients obtained by eCMMcv.

### 3.1.2 Dual avoided crossing

In the DAC model, the diagonal elements of the potential operator are $V_{11}=0$ and $V_{22}=-Ae^{-BR^2}+E_0$, and the off-diagonal ones are $V_{12}=V_{21}=Ce^{-DR^2}$, where the parameters are set as $A=0.10$, $B=0.28$, $E_0=0.05$, $C=0.015$ and $D=0.06$. Two crossing points appear in the diagonal potential energy surfaces. At time $t=0$ state $I$ is occupied with the initial nuclear wavepacket $\Psi(R;t=0) \propto \exp[-\alpha(R-R_0)^2/2 + i(R-R_0)P_0/\hbar]$, where the Gaussian width parameter is $\alpha=1$ a. u., the initial average coordinate is $R_0=-10$ a. u., and initial average momentum $P_0$ varies from 2 a. u. to 50 a. u..

Results for the transmission coefficients are presented in Figure 3 and Figure 4. Panels 3(a) and 3(b) show the eCMM results on state 1 and on state 2, respectively, using the diabatic representation. The results produced by Ehrenfest (mean field) dynamics demonstrate a noticeable deviation from the exact DVR data for $P_0 \geq 10$ a. u.. The eCMM approach with either $\gamma=\left(\sqrt{F+1}-1\right)/F=0.366$ or $\gamma=1/2$ yields accurate results for the relatively large initial average momentum ($P_0 \geq 10$ a. u.). It captures the correct shape of Stückelberg oscillations[14, 57, 96]. The eCMM results calculated in the adiabatic representation shown in Panels 3(c) and 3(d) are consistent with those in the diabatic representation demonstrated in Panels 3(a) and 3(b). Figure 4 presents the eCMMcv results for the same model. In comparison to the eCMM results of Figure 3, the eCMMcv data of Figure 4 show better performance in a large range of initial average momentum and are also less sensitive to parameter $\gamma$.

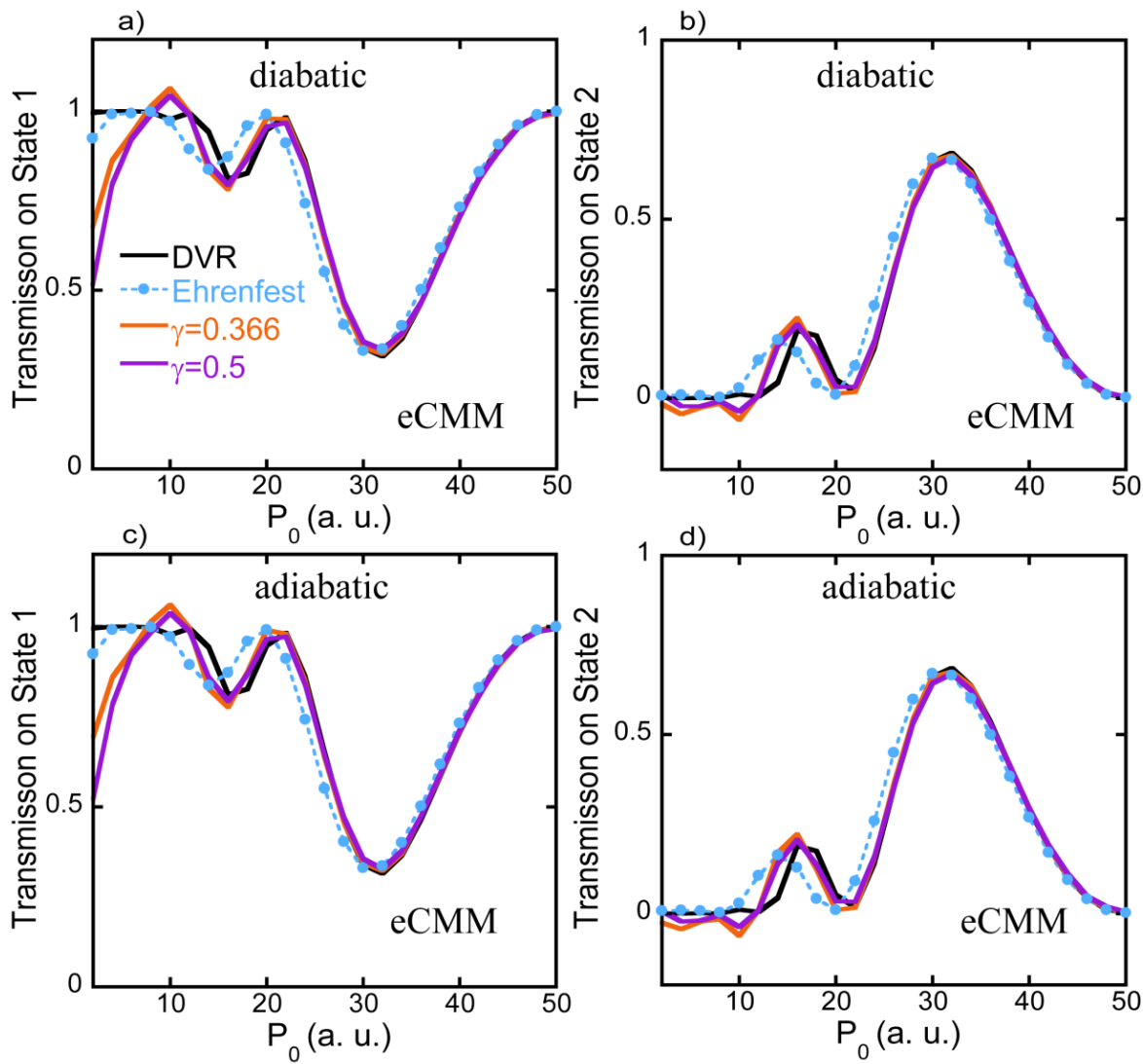

**Figure 3:** Transmission coefficients of Tully's DAC model. In panel a, orange and purple solid lines: eCMM transmission coefficients on state 1 for $\gamma = 0.366$ and $\gamma = 0.5$, respectively. Black solid line: Exact results yielded by DVR. Blue dashed line with circles: Ehrenfest dynamics. Panel b is similar to panel a but for the eCMM transmission coefficients on state 2. Panels c and d are the same as panels a and b, respectively, except that the adiabatic representation is used in panels

c and d. The transmission coefficients in the adiabatic representation are transformed to those in the diabatic representation for fair comparison.

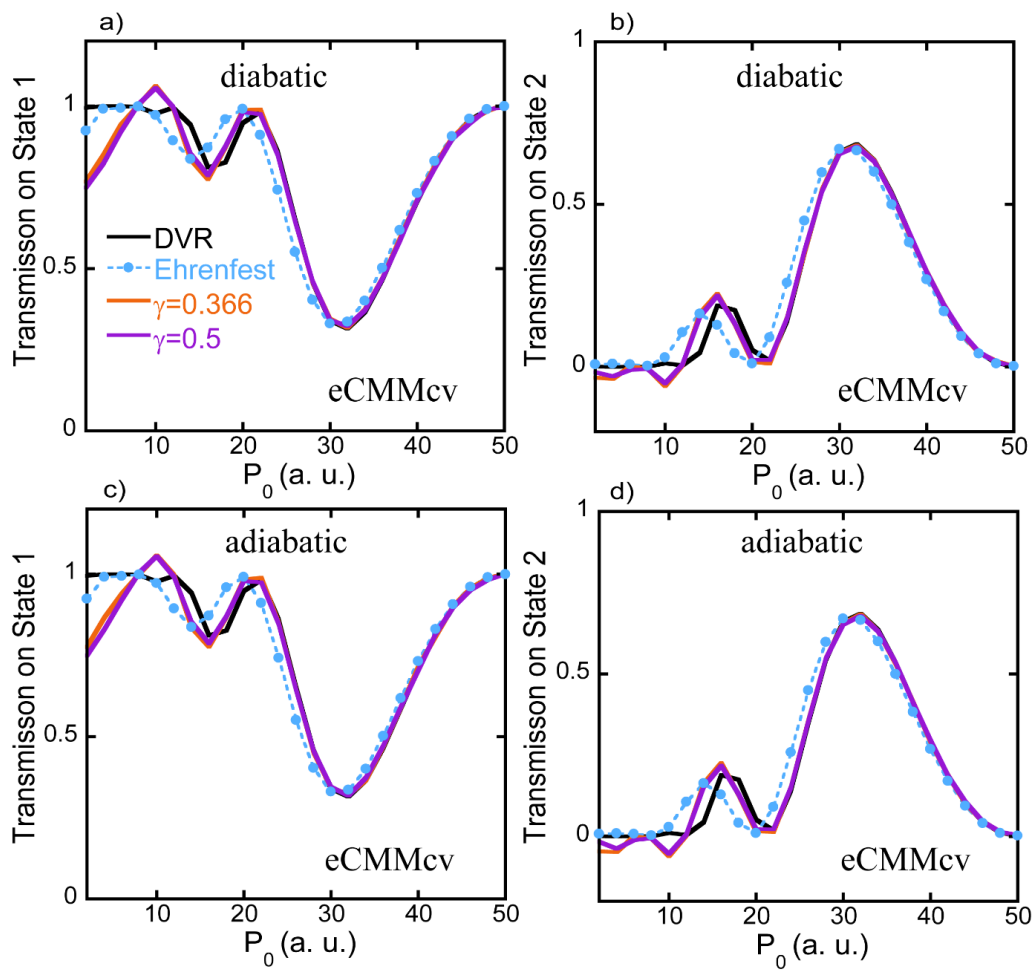

**Figure 4:** Same as Figure 3, but for transmission coefficients obtained by eCMMcv.

### 3.2. Three-state photodissociation models of Miller and coworkers

The coupled three states with Morse oscillators proposed by Miller and coworkers[58], which mimic ultrafast photo-dissociation processes, provide another set of gas phase benchmark models for testing nonadiabatic dynamics methods. The models are composed of three Morse potentials with Gaussian coupling terms, which are of the form

$$V_{ii}(x) = D_i \left[1 - e^{-\beta_i(x-R_i)}\right]^2 + C_i, \quad i = 1, 2, 3.$$
$$V_{ij}(x) = V_{ji}(x) = A_{ij} e^{-\alpha_{ij}(x-R_{ij})^2}, \quad i, j = 1, 2, 3; \text{ and } i \neq j.$$
(45)

The parameters for the three models are listed in Table 1. The nuclear mass is set to $m = 20000$ a.u.. The initial Gaussian wavepacket for the nuclear DOF is prepared from a ground state with frequency $\omega = 5 \times 10^{-3}$ au $= 1097$ cm$^{-1}$ centering in $R_e = 2.9, 3.3, 2.1$ a.u. for Models 1, 2 and 3, respectively. The Wigner distribution for the nuclear DOF is

$$\rho(x, p) \propto \exp\left[-m\omega(x - R_e)^2 / \hbar - p^2 / (m\omega\hbar)\right] \quad . \quad (46)$$

We use 96,000 trajectories to yield fully converged data in each eCMM or eCMMcv simulation.

The three systems have been studied by a few nonadiabatic dynamics methods based on the Meyer-Miller mapping Hamiltonian[29, 58, 90, 94]. Because the Gaussian coupling terms of eq 45 are relatively local, short-time dynamics implies the Born-Oppenheimer limit.

Table 1： Parameters of 3-State Photodissociation Morse potential models [58]

| Parameters | Model 1 | Model 2 | Model 3 |
|---|---|---|---|
| $D_1, D_2, D_3$ | 0.003, 0.004, 0.003 | 0.020, 0.010, 0.003 | 0.020, 0.020, 0.003 |
| $\beta_1, \beta_2, \beta_3$ | 0.65, 0.60, 0.65 | 0.65, 0.40, 0.65 | 0.40, 0.65, 0.65 |
| $R_1, R_2, R_3$ | 5.0, 4.0, 6.0 | 4.5, 4.0, 4.4 | 4.0, 4.5, 6.0 |
| $C_1, C_2, C_3$ | 0.00, 0.01, 0.006 | 0.00, 0.01, 0.02 | 0.02, 0.00, 0.02 |
| $A_{12}, A_{23}, A_{31}$ | 0.002, 0.002, 0.0 | 0.005, 0.0, 0.005 | 0.005, 0.0, 0.005 |

| $R_{12}, R_{23}, R_{31}$ | 3.40, 4.80, 0.00 | 3.66, 0.00, 3.34 | 3.40, 0.00, 4.97 |
| $\alpha_{12}, \alpha_{23}, \alpha_{31}$ | 16.0, 16.0, 0.0 | 32.0, 0.0, 32.0 | 32.0, 0.0, 32.0 |

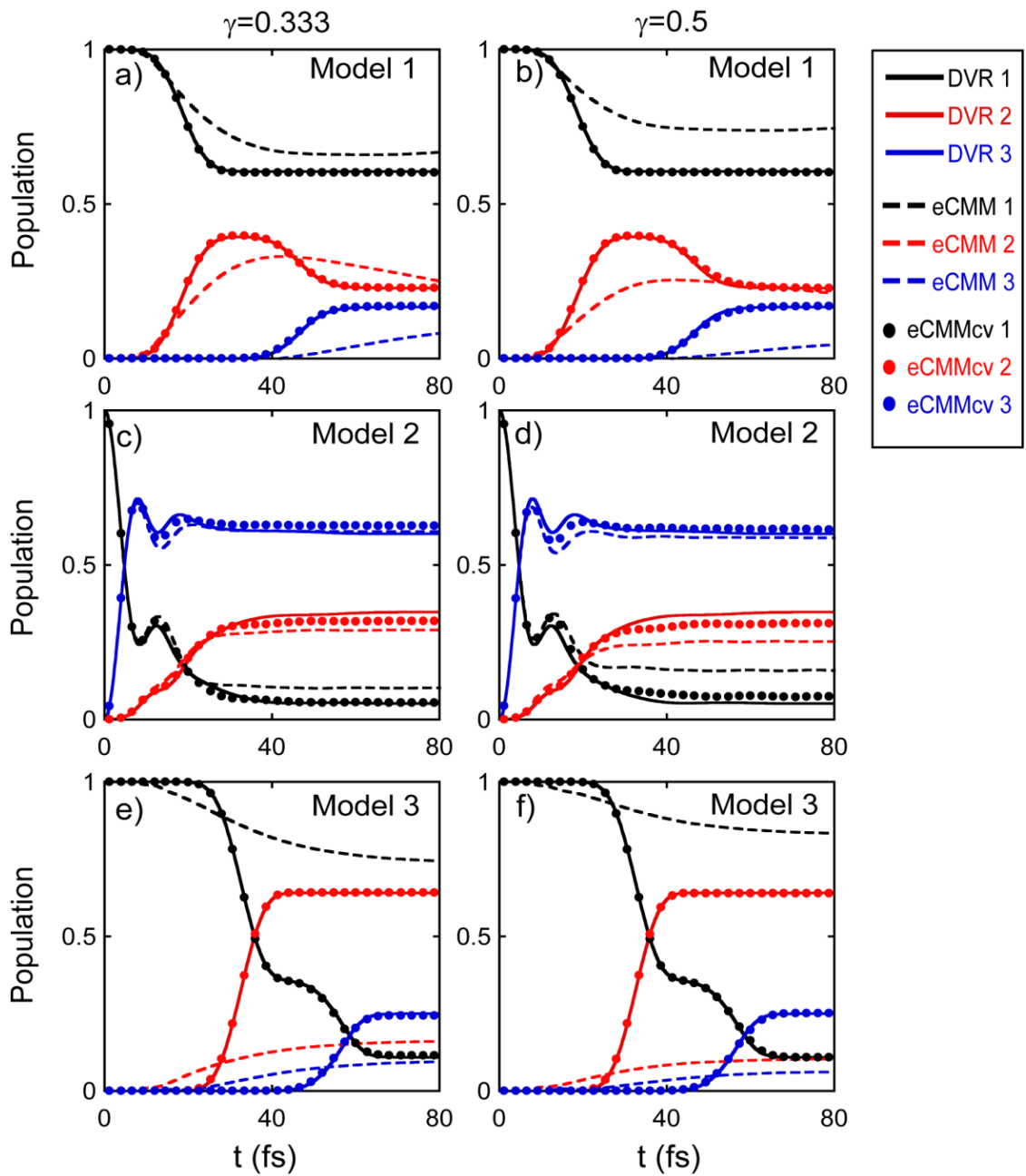

**Figure 5:** Population dynamics results of the three states for photo-dissociation models[58] listed in Table 1. The eCMM as well as eCMMcv results are obtained with $\gamma = 0.333$ or $\gamma = 0.5$. Black, Red and Blue markers: Populations on state 1, state 2 and state 3, respectively. Solid lines: Exact results produced by DVR. Dashed lines: The eCMM results. Solid circles: The eCMMcv results. Panels (a) and (b) for *Model 1*, Panels (c) and (d) for *Model 2*, and Panels (e) and (f) for *Model 3*.

Figure 5 compares the exact population of each state in each model as a function of time produced by DVR, that generated by eCMM, and that yielded by eCMMcv. While Panels 5(a), 5(c) and 5(e) demonstrate the eCMM/eCMMcv results with parameter $\gamma = \left(\sqrt{F+1} - 1\right)/F = 0.333$ for Models 1, 2 and 3, respectively, Panels 5(b), 5(d) and 5(f) show such results with $\gamma = 1/2$. Figure 5 indicates that eCMMcv is overall superior to eCMM in the three model tests. The eCMMcv approach is more accurate as well as less sensitive to parameter $\gamma$. This is mainly because eCMMcv approaches the Born-Oppenheimer limit when the state-state coupling disappears at short times.

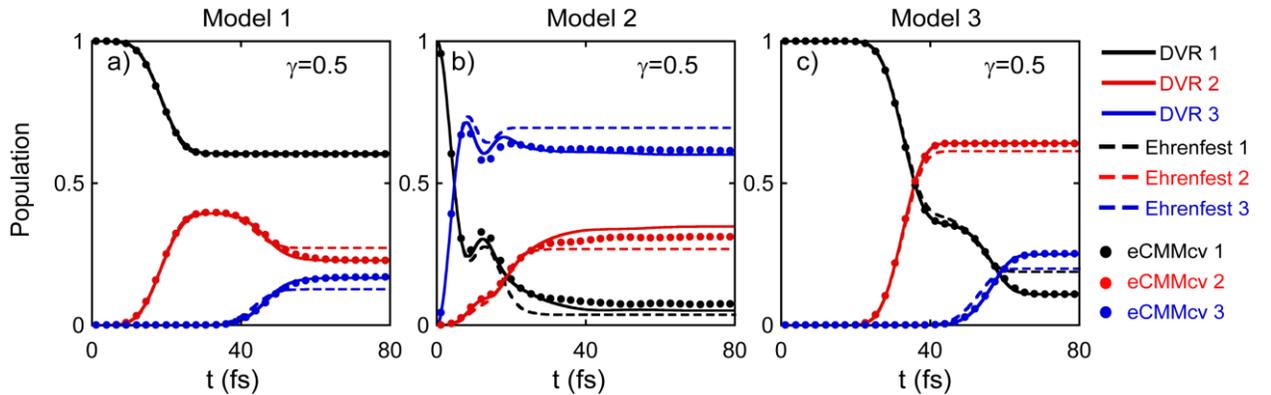

**Figure 6:** Comparison between eCMMcv and Ehrenfest dynamics. Solid circles: The eCMMcv results with $\gamma = 0.5$. Solid lines: Exact results generated by DVR. Dashed lines: Ehrenfest dynamics. Black, Red, Blue colors represent populations on the first, second and third state, respectively. Panels (a), (b), (c) are for *Model 1, 2* and *3*, respectively.

In Figure 6 we compare the performance of eCMMcv (with $\gamma = 1/2$ for demonstration) to that of Ehrenfest dynamics. Ehrenfest dynamics produces reasonable short time results but yields significant deviation from the long time limit. In comparison, eCMMcv generates much more accurate results for these three models.

### 3.3. FMO monomer

The Fenna-Matthews-Olson (FMO) monomer is a benchmark site-exciton or system-bath model widely used for testing nonadiabatic dynamics methods[25, 28, 37, 44, 46, 49, 59, 97-105]. The FMO monomer model includes seven sites, and each site denotes a photosynthetic pigment (Bacteriochlorophyll). The 7-site system is described by the Hamiltonian of Ref. [59] (in units of wavenumber),

$$H_s = \begin{pmatrix} 12410 & -87.7 & 5.5 & -5.9 & 6.7 & -13.7 & -9.9 \\ -87.7 & 12530 & 30.8 & 8.2 & 0.7 & 11.8 & 4.3 \\ 5.5 & 30.8 & 12210 & -53.5 & -2.2 & -9.6 & 6.0 \\ -5.9 & 8.2 & -53.5 & 12320 & -70.7 & -17.0 & -63.3 \\ 6.7 & 0.7 & -2.2 & -70.7 & 12480 & 81.1 & -1.3 \\ -13.7 & 11.8 & -9.6 & -17.0 & 81.1 & 12630 & 39.7 \\ -9.9 & 4.3 & 6.0 & -63.3 & -1.3 & 39.7 & 12440 \end{pmatrix}, \quad (47)$$

and surrounding protein environments are depicted by harmonic baths, $H_b = \sum_{n,i} \frac{1}{2}\left(P_{ni}^2 + \omega_{ni}^2 R_{ni}^2\right)$, where $\{R_{ni}, P_{ni}, \omega_{ni}\}$ are the position, momentum and frequency for the $i$-th bath mode on site $n$, respectively. Interaction between the system and bath modes adopts a bilinear form, $H_{sb} = -\sum_{n,i} c_{ni} R_{ni} |n\rangle\langle n|$ with $c_{ni}$ the exciton-phonon or system-bath coupling coefficient, which can be determined from the discretization of the spectral density of the bath. The bath is characterized by a Debye spectral density[83, 106, 107], which adopts a Lorentzian cutoff,

$$J(\omega) = 2\lambda \frac{\omega_c \omega}{\omega_c^2 + \omega^2} \ , \tag{48}$$

where $\lambda$ is the bath reorganization energy and $\omega_c$ is the characteristic frequency. A proper discretization scheme is[108],

$$\begin{cases} \omega_{ni} = \omega_c \tan\left[\pi/2\left(1 - i/(1+N_b)\right)\right] \\ c_{ni} = \omega_{ni}\sqrt{2\lambda/(1+N_b)} \end{cases}, \quad i = 1,\cdots,N_b \ . \tag{49}$$

Here $N_b$ is the total number of discretized harmonic modes and we employ $N_b = 50$ modes per site for converged results. The bath parameters are $\{\lambda = 35 \text{ cm}^{-1}, \omega_c = 106.14 \text{ cm}^{-1}\}$. We study a relatively low temperature $T = 77$ K, which is a challenging case for many nonadiabatic dynamics methods. We consider two different cases. In the first case the initial excitation occurs on pigment/site 1, and in another case pigment/site 6, instead, is excited at the beginning. Since coherence effects could be important in the photo-harvesting system, we calculate the population of each site as well as the electronic coherence terms (i.e., the off-diagonal elements of the reduced density matrix for the electronic DOFs). An ensemble of 120,000 trajectories are used in each eCMM or eCMMcv simulation. Several numerically exact approaches are capable of offering benchmark results for the FMO monomer model, which include quasi-adiabatic propagator path integral (QuAPI)[109-111], hierarchical equations of motion (HEOM)[112-120] and multi-layer multi-configurational time-dependent Hartree (ML-MCTDH)[107, 121-125]. We utilize HEOM to obtain exact results for the FMO model system. Parameter $\gamma = \left(\sqrt{F+1}-1\right)/F \approx 0.261$ or $\gamma = 1/2$ is used for eCMM and eCMMcv.

Figure 7 shows the population of each site of the FMO monomer when site *1* is initially excited. While Panels 7(a) and 7(b) demonstrate the eCMM results for $\gamma = \left(\sqrt{F+1}-1\right)/F \approx 0.261$ and $\gamma = 1/2$, respectively, Panels 7(c) and 7(d) present the corresponding results generated by eCMMcv. Panel 7(e) shows that Ehrenfest dynamics works poorly in this case. Figure 8 demonstrates the same information as Figure 7, but for the initial excitation on site 6 instead. It is indicated in Figure 7 and Figure 8 that the results yielded by eCMM are close to those by eCMMcv, which are reasonably accurate in comparison to exact data. The eCMMcv approach performs slightly better than eCMM for this site-exciton model system.

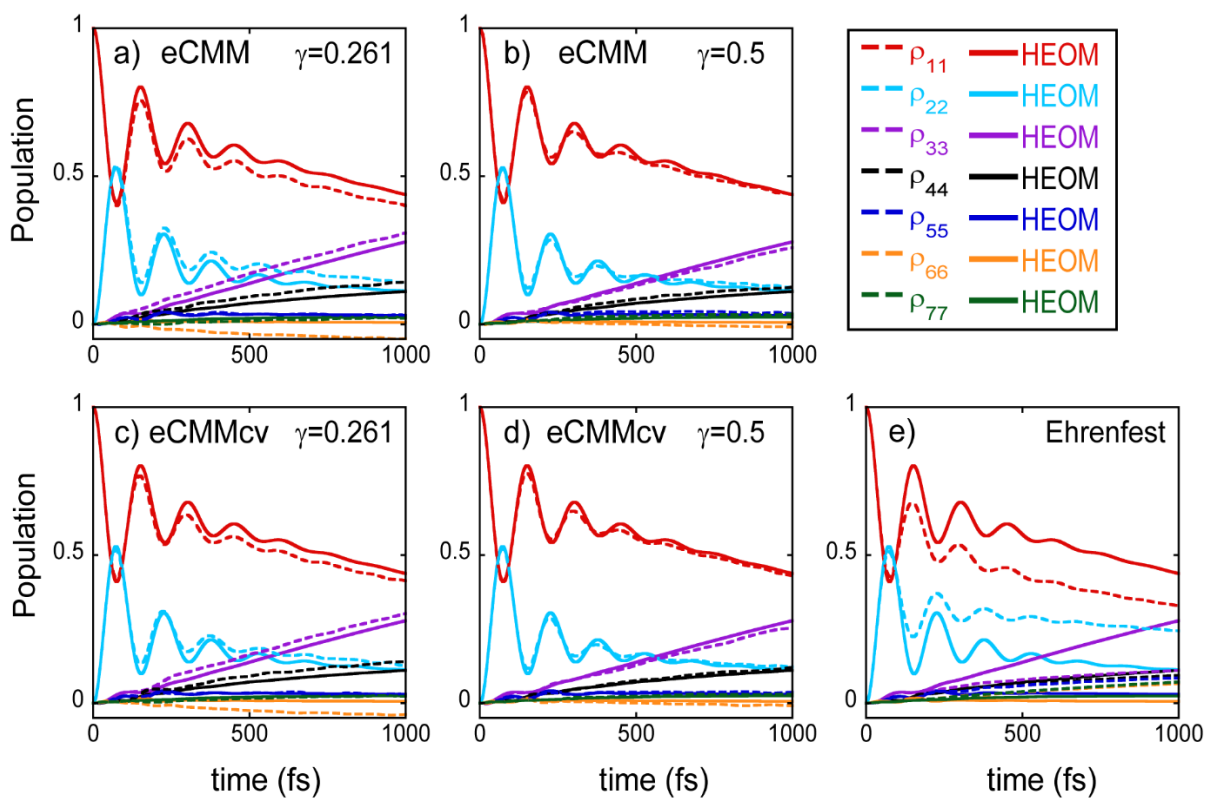

**Figure 7:** Population dynamics of the 7-state site-exciton model for FMO at 77K, where the initial excitation is on the first pigment (Site 1). Panels (a) and (b) show the eCMM results with parameter $\gamma = 0.261$ and those with $\gamma = 0.5$, respectively. Red, Blue, Purple, Black, Blue, Orange and Green lines present populations of site 1,2,3,4,5,6, and 7, respectively. Dashed lines: The eCMM results. Solid lines: Exact results by HEOM. Panels (c) and (d) are same as Panels (a) and (b), respectively, but for the eCMMcv results. In Panel (e), dashed lines are used for Ehrenfest dynamics results, while solid lines are for HEOM results.

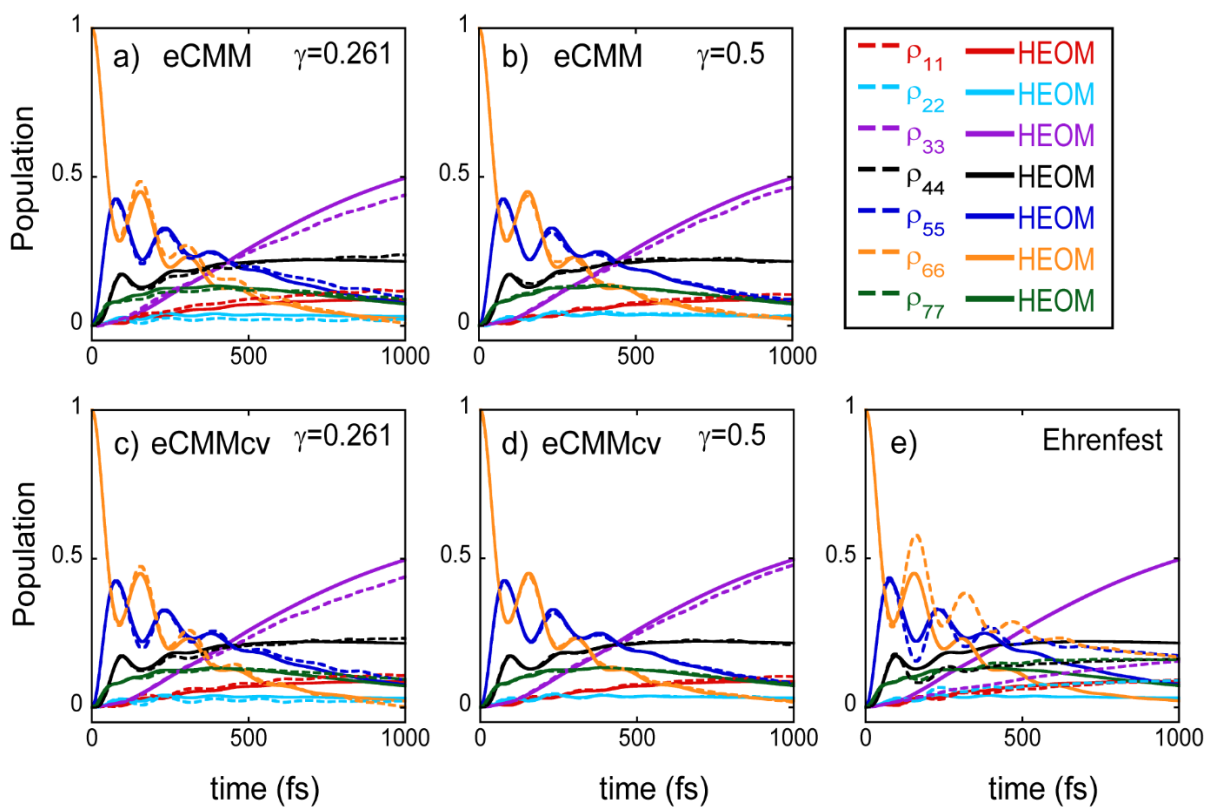

**Figure 8:** Same as Figure 7, but for the case where the initial excitation occurs on the sixth pigment (Site 6).

We then study the (electronic) coherence terms. The four most important off-diagonal elements (which have the largest absolute values) of the reduced density matrix are selected for demonstration. The moduli of $\rho_{12}$, $\rho_{13}$, $\rho_{15}$, and $\rho_{34}$ are illustrated in Figure 9 for the case where site 1 is initially excited. When site 6 is excited at the beginning, the moduli of $\rho_{34}$, $\rho_{45}$, $\rho_{47}$, and $\rho_{56}$ are presented in Figure 10. In comparison to the poor performance of Ehrenfest dynamics [as shown in Panel 9(e) or Panel 10(e)], either of eCMM and eCMMcv yields much more reasonably good results. The eCMMcv results are slightly closer to the HEOM data than the eCMM ones in Figure 9 and Figure 10.

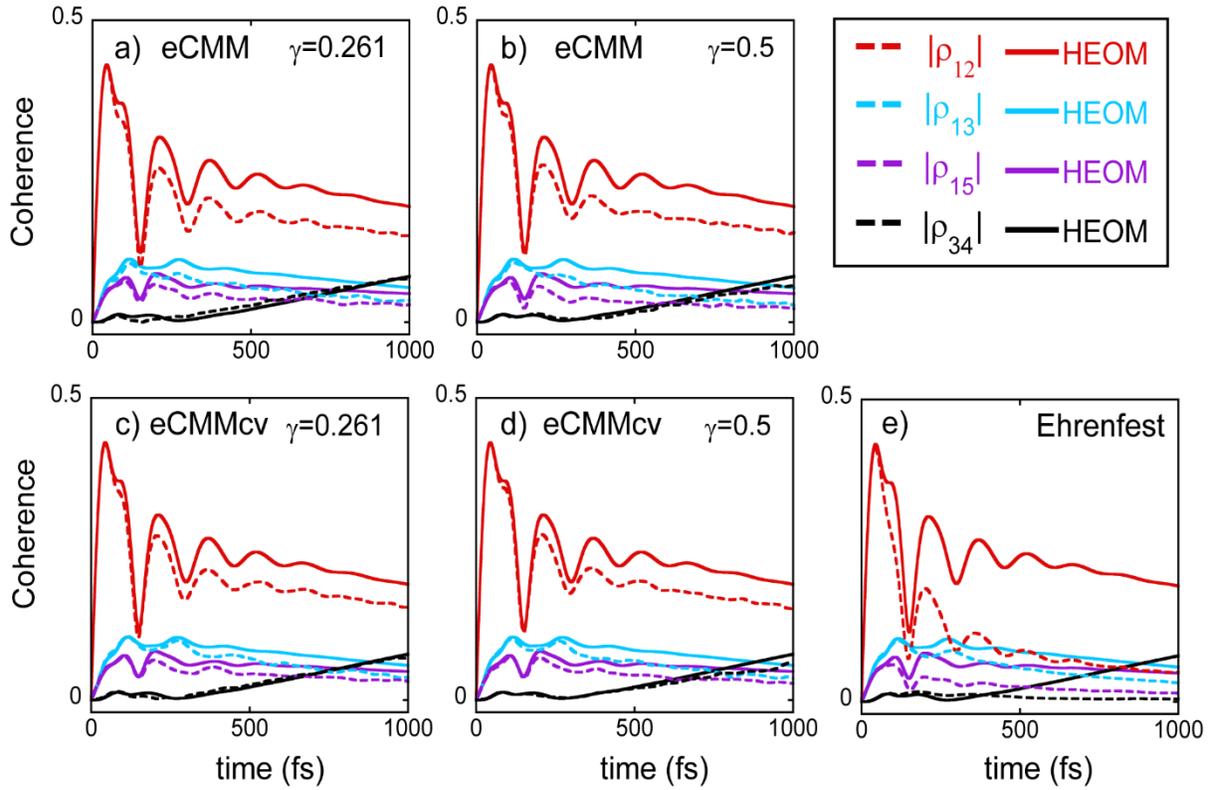

**Figure 9:** Coherence terms of the 7-state site-exciton model for FMO at 77K, where the initial excitation is on the first pigment (Site 1). Panels (a) and (b) show the eCMM results with parameter $\gamma = 0.261$ and those with $\gamma = 0.5$, respectively. Red, Blue, Purple, and Black colors are used for $|\rho_{12}|$, $|\rho_{13}|$, $|\rho_{15}|$ and $|\rho_{34}|$, respectively. Dashed lines: The eCMM results. Solid lines: Exact results by HEOM. Panels (c) and (d) are same as Panels (a) and (b), respectively, but for the eCMMcv results. In Panel (e), dashed lines are used for Ehrenfest dynamics results, while solid lines are for HEOM results.

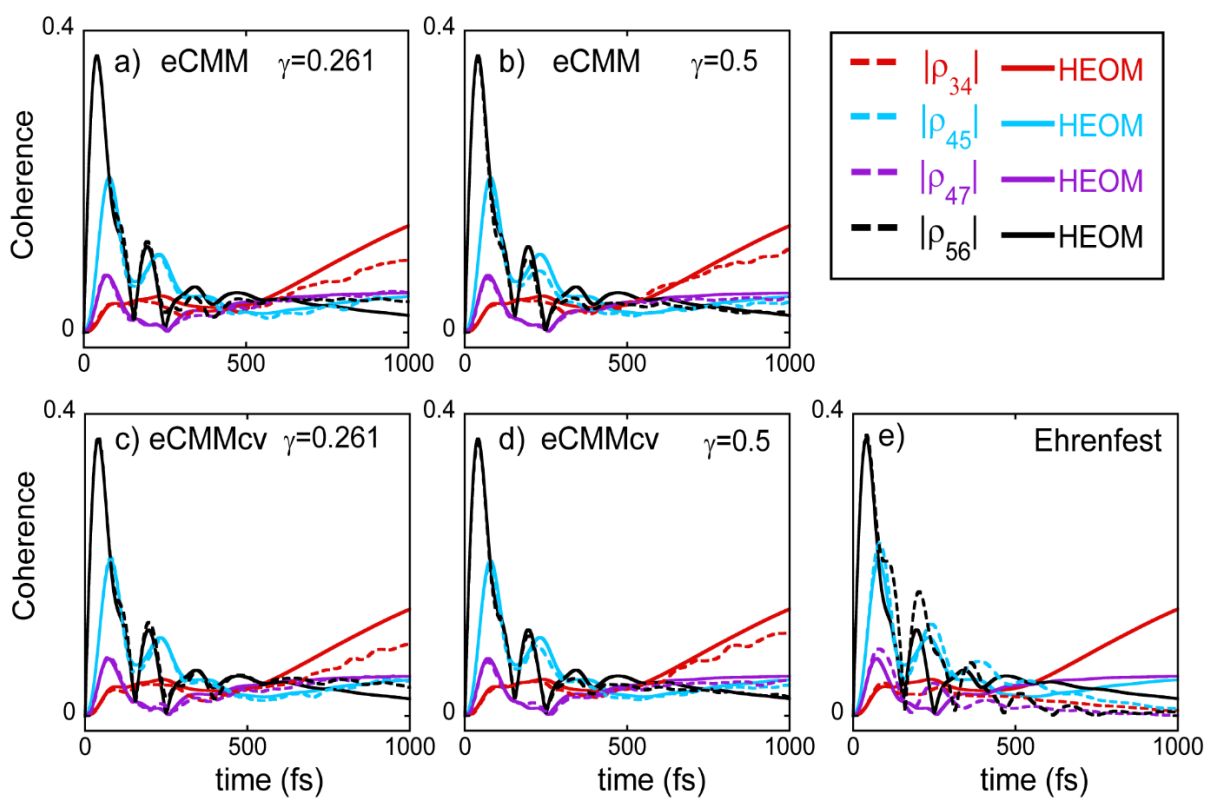

**Figure 10:** Same as **Figure 9**, but for the case where the initial excitation occurs on the sixth pigment (Site 6). In this case, Red, Blue, Purple, and Black lines are used to present $|\rho_{34}|$, $|\rho_{45}|$, $|\rho_{47}|$ and $|\rho_{56}|$, respectively.

More results on the population dynamics of each site of the FMO monomer for 77K and 300K are available in **Section S3** of Supporting Information, in which the eCMM/eCMMcv approaches are compared to Ehrenfest dynamics as well as HEOM. Provided that the eCMM/eCMMcv approaches lead to overall satisfying short-time as well as long-time dynamics results for the FMO monomer for 77K and 300K, it is reasonable to expect that eCMM/eCMMcv can in principle predict semi-quantitative data for very low temperature or even 0K, where it is generally difficult for HEOM to obtain converged data. The reliable performance of eCMM for the spin-boson model at 0K has already been demonstrated in Refs. [41, 42]. Figure 11 compares the population dynamics for site 1 as well as site 3 for the FMO monomer at different temperatures when site 1 is initially activated. As the temperature decreases, the relaxation time scale increases. While the oscillating behavior (of the population dynamics of site 1) vanishes after only two periods (less than 300 fs) at 300K, such behavior lasts significantly longer than 1000 fs at 0K as shown in Figure 11.

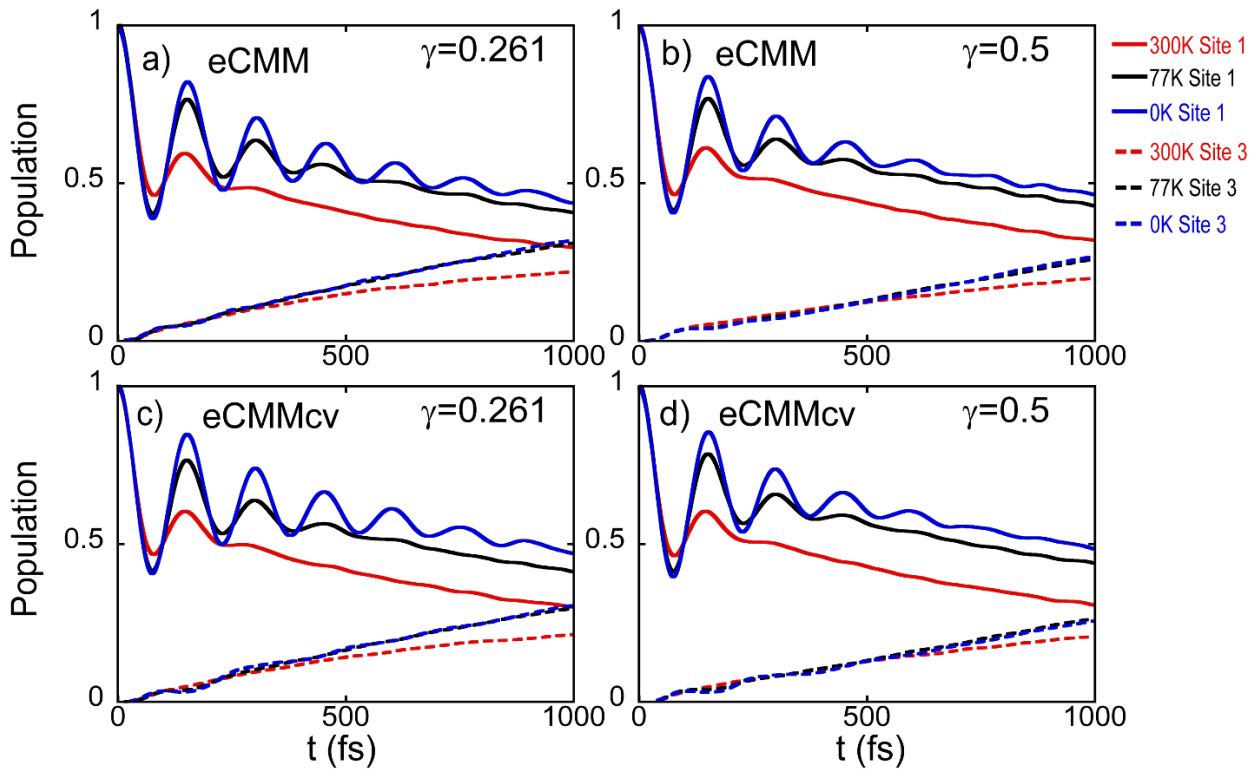

**Figure 11**. Dynamics at different temperatures for the FMO monomer model when site 1 is initially activated. Red, Black, Blue lines represent populations at 300K, 77K, and 0K, respectively. Solid

lines: The eCMM/eCMMcv results for site 1 (pigment 1). Dashed lines: The eCMM/eCMMcv results for site 3 (pigment 3).

### 3.4. Atom-in-cavity models

Mixed quantum-classical trajectory-based methods have been utilized to study interaction dynamics of light and matter, which offer approximate but practical approaches for simulating realistic systems in chemistry, materials, and biology[126-136]. We utilize eCMM and eCMMcv to test the performance of describing the cavity-modified chemical dynamics. The benchmark system that we study in this section involves interaction between an atom with frozen nuclear DOFs in a one-dimmensional lossless cavity, which exhibits relaxation dynamics (for the atom) in conjunction with spontaneous emission of photons.[50, 60-63, 93, 134]. The atomic electronic states are coupled because of the interaction of the cavity field and the transition moments between different atomic energy levels. Such a system can be elaborated as a multi-electronic-state Hamiltonian model, of which the off-diagonal terms are from the coupling with the cavity field. After making the dipole approximation[137, 138] and neglecting the second order interaction (which only leads to a constant shift of the energy level in a two-state system[139]), a general Hamiltonian reads[62, 63],

$$\hat{H} = \sum_{k=1}^{N_e} \varepsilon_k |k\rangle\langle k| + \sum_{k,k'}^{N_e} \sum_{\alpha=1}^{N_p} \omega_\alpha \hat{R}_\alpha \lambda_\alpha \mu_{kk'} |k\rangle\langle k'| + \frac{1}{2}\sum_{\alpha=1}^{N_p} \left( \hat{P}_\alpha^2 + \omega_\alpha^2 \hat{R}_\alpha^2 \right) \quad , \qquad (50)$$

where $|k\rangle$ is the $k$-th atomic energy level, $\{\hat{P}_\alpha, \hat{R}_\alpha, \omega_\alpha\}$ denote the momenta, positions and frequencies of the photonic modes in the cavity, $N_e$ and $N_p$ stand for the number of atomic energy levels and that of cavity field modes, respectively, $\mu_{kk'}$ is the transition dipole moment between two energy levels, $|k\rangle$ and $|k'\rangle$, and $\lambda_\alpha \equiv \lambda_\alpha(r_A)$ represents the coupling between the $\alpha$-th field mode and the atom at fixed position $r_A$. The values of the parameters of the Hamiltonian are given in Refs. [50, 62, 63], which we briefly describe below. (All parameters are used in atomic units.)

The cavity mode frequencies are determined by the standing-wave condition, i.e.,

$$\omega_\alpha = \frac{\alpha c \pi}{L}, \quad \alpha = 1, \cdots, N_p, \quad (51)$$

here $N_p = 400$ and the coupling vector $\lambda_\alpha(r_A)$ with fixed atom position $r_A$ is

$$\lambda_\alpha(r_A) = \sqrt{\frac{2}{\varepsilon_0 L}} \sin\left(\frac{\alpha \pi r_A}{L}\right), \quad (52)$$

where $c = 137.036$ stands for the speed of light, $\varepsilon_0 = 1/(4\pi)$ represents the vacuum permittivity, $L = 2.362 \times 10^5$ is the volume length of the cavity, and the atom is fixed at the center of the cavity $r_A = L/2$. Two models are considered: one is a two-level atom model with atomic energy levels $\varepsilon_1 = -0.6738, \varepsilon_2 = -0.2798$ and transition dipole moment $\mu_{12} = 1.034$, the other is a three-level atom model with atomic energy levels $\varepsilon_1 = -0.6738, \varepsilon_2 = -0.2798, \varepsilon_3 = -0.1547$, and transition dipole moments $\mu_{12} = 1.034$, $\mu_{23} = -2.536$. Four hundred field modes are used in each model. We choose the highest excited state of the atom as the initial state. Each cavity mode is initially in its vacuum state $\langle R_\alpha | \Psi_0 \rangle \propto \exp[-\omega_\alpha R_\alpha^2/(2\hbar)]$ (with zero number of photons). The Wigner distribution of the initial density operator for the cavity field modes then reads

$$\rho_W(\mathbf{R}, \mathbf{P}) \propto \prod_{\alpha=1}^{N_p} \exp\left[-\left(P_\alpha^2/(\hbar \omega_\alpha) + \omega_\alpha R_\alpha^2/\hbar\right)\right]. \quad (53)$$

An ensemble of 96,000 trajectories are used to yield fully converged results for eCMM/eCMMcv. Exact results for the cavity quantum electrodynamics processes in the two models can be obtained by the truncated configuration interaction (CI) approach, which are available in Refs. [62, 63].

Figure 12 and Figure 13 show population dynamics of the two-level model and that of the three-level model, respectively. The eCMM and eCMMcv results are obtained with $\gamma = \left(\sqrt{F+1}-1\right)/F = 0.366$ or $\gamma = 1/2$ for the 2-level model, and with $\gamma = \left(\sqrt{F+1}-1\right)/F = 0.333$ or $\gamma = 1/2$ for the 3-level model. The exact data[62, 63] and Ehrenfest dynamics results are also demonstrated for comparison. As shown in Panel 12(e) and Panel 13(e), the Ehrenfest dynamics results considerably deviate from the exact population dynamics even at very short times, which agrees with what was reported in Refs. [50, 62, 63]. In contrast, Ehrenfest dynamics yields reasonable behavior at very short times in the previous benchmark models studied in the paper, although its long time performance is often poor. This indicates that the atom-in-cavity models are even more challenging for testing the outperformance of a nonadiabatic dynamics method beyond Ehrenfest dynamics. It has been demonstrated in Refs. [62, 63] that the fewest switches surface hopping approach[57] generates even worse results than Ehrenfest dynamics.

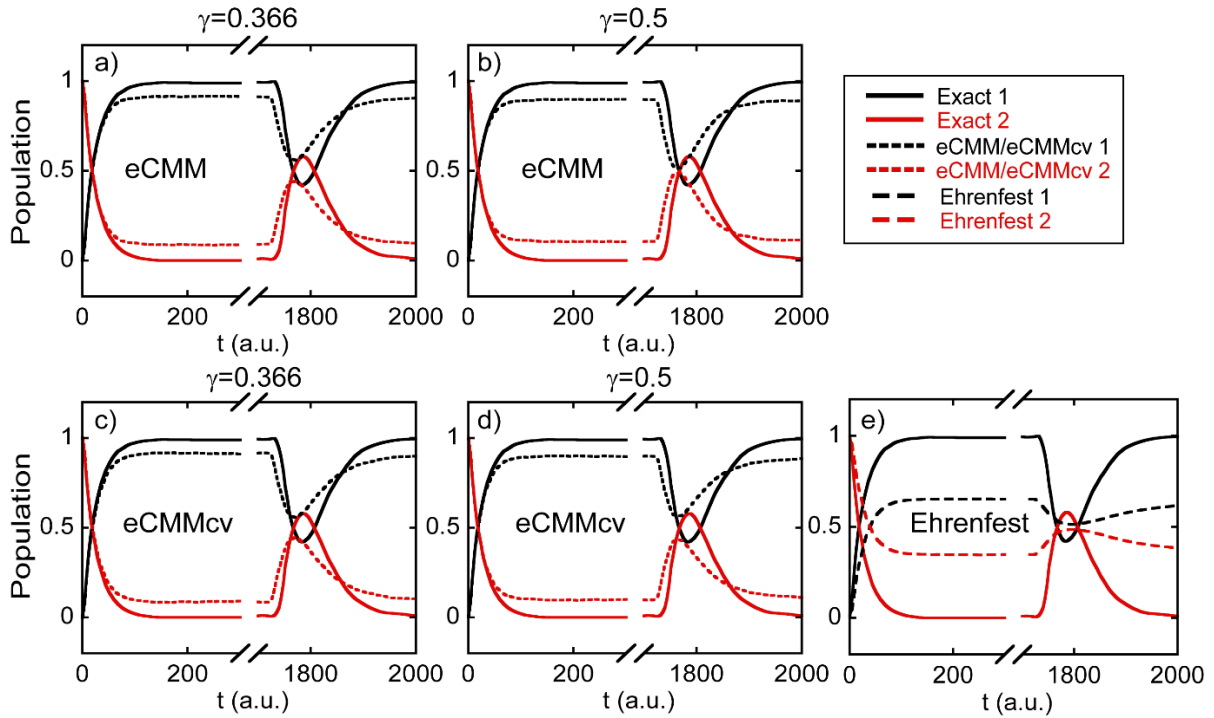

**Figure 12.** Population as a function of time for the two-level atomic model in optical cavity. Panel (a) shows the eCMM results with $\gamma = 0.366$, while Panel (b) demonstrates those with $\gamma = 0.5$. Panels (c) and (d) are same as Panels (a) and (b), respectively, but for the eCMMcv results. Ehrenfest dynamics results are presented in Panel (e) for comparison. Black color: Population of State 1. Red color: Population of State 2. Solid lines: Exact results from Refs. [62, 63]. Short-dashed lines: eCMM results in Panels (a) and (b), or eCMMcv results in Panels (c) and (d). Long-dashed lines: Ehrenfest dynamics results in Panel (e).

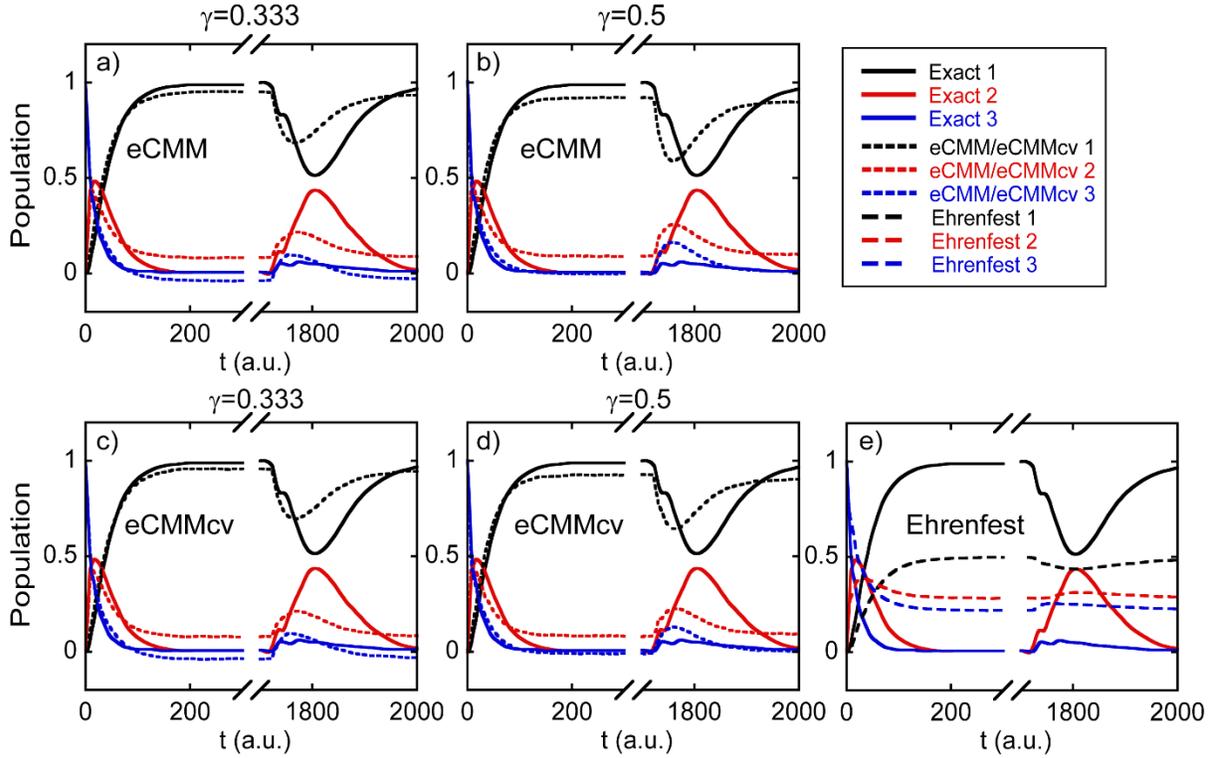

**Figure 13.** Population as a function of time for the three-level atomic model in optical cavity. Panel (a) shows the eCMM results with $\gamma = 0.333$, while Panel (b) demonstrates those with $\gamma = 0.5$. Panels (c) and (d) are same as Panels (a) and (b), respectively, but for the eCMMcv results. Ehrenfest dynamics results are presented in Panel (e) for comparison. Black, Red, and Blue colors: Population of State 1, that of State 2, and that of State 3. Solid lines: Exact results from Refs. [62, 63]. Short-dashed lines: eCMM results in Panels (a) and (b), or eCMMcv results in Panels (c) and (d). Long-dashed lines: Ehrenfest dynamics results in Panel (e).

Figure 12(a)-(d) and Figure 13(a)-(d) demonstrate that the results yielded by eCMMcv are very similar to those produced by eCMM. Either of eCMM or eCMMcv achieves significantly better performance than Ehrenfest dynamics in describing the cavity-modified chemical dynamics of the two models. The re-absorption and re-emission process of the earlier emitted photon by the atom occurs around $t = 1800$ a.u. It is encouraging that both eCMM and eCMMcv are capable of semi-quantitatively depicting the positive (negative) spike in the excited (ground) electronic state of the atom. Figures 12-13 can be compared to Figures 8 and 12 of Ref. [63] as well as Figures 1-2 of Ref. [50].

Although eCMMcv is developed in an exact phase space mapping formulation of the correlation function as described in Section 2.1, its trajectory-based dynamics governed by the general mapping Hamiltonian with commutator variables (eq 29 or eq 36) is nevertheless an approximation to the exact equations of motion in quantum mechanics. Although the eCMMcv approach leads to overall reasonably good results in various model tests as shown in Figures 1-13, in the future it will be useful to overcome several drawbacks of eCMMcv. As shown in Figures 3-4, neither eCMM nor eCMMcv performs well in the low momentum region, which implies that the performance of eCMMcv should be improved in order to faithfully describe the deep tunneling regime. Figures 3-4, Figures 7-8, and Figure 13 indicate that negative values for the population of a site/state can occasionally occur in eCMMcv results, while the window function treatment of the SQC approach[29] is expected to solve such a problem. Like most approximate nonadiabatic dynamics methods, the eCMMcv approach does not guarantee that the detail balance is rigorously satisfied for both electronic and nuclear DOFs when the whole system is at thermal equilibrium, although the long time limit results for the electronic DOFs for most model tests are reasonably good. It will be interesting to see how the strategies of Refs. [140-142] and of Refs. [143, 144] can practically be used to systematically improve the mapping Hamiltonian dynamics for multi-dimensional nonadiabatic systems.

## 4. Conclusion remarks

In the conceptually different picture presented in the unified framework for phase space mapping models[13], it is indicated that there exists a more comprehensive mapping Hamiltonian (eq 29 or eq 36) beyond the well-known Meyer-Miller Hamiltonian[11, 12] (eq 3), where commutator matrix $\mathbf{\Gamma}$ that consists of auxiliary mapping variables for $-\left\{\frac{i}{8}\left(\left[\hat{\sigma}_x^{(n)},\hat{\sigma}_y^{(m)}\right]+\left[\hat{\sigma}_x^{(m)},\hat{\sigma}_y^{(n)}\right]\right)\right\}$, rather than the conventional zero-point-energy parameter, is involved. In the exact mapping formulation on constraint space for phase space approaches for nonadiabatic dynamics[41, 42], such a general mapping Hamiltonian with commutator variables (eq 29 or eq 36) can be used to produce eCMMcv, an approximate trajectory-based approach. We have tested a few benchmark models that range from gas phase to condensed phase systems, which include the SAC and DAC scattering models[57], 3-state photodissociation models[58], 7-site model of the Fenna-Matthews-Olson (FMO) monomer[59], and atom-in-cavity models[60-63]. Parameter $\gamma$ in the exact mapping kernel is recommended in the region, $\left[\left(\sqrt{F+1}-1\right)/F,\ 1/2\right]$, where the eCMMcv results are relatively insensitive to the value of $\gamma$. The results demonstrate that the overall performance of the general mapping Hamiltonian (eq 29 or eq 36) employed in eCMMcv is better than the original Meyer-Miller Hamiltonian (eq 3) used in eCMM.

The conclusion applies to the most recent version of symmetrical quasi-classical (SQC) dynamics with triangle window functions[29]. The successful SQC methods of Cotton and Miller employ the conventional Meyer-Miller Hamiltonian or its symmetrized form[18-29]. As shown in **Section S4** of Supporting Information, when the original Meyer-Miller Hamiltonian is replaced by the general mapping Hamiltonian with commutator variables (eq 29 or eq 36), the performance of the latest SQC approach of Ref. [29] can be improved. It is expected that the general mapping Hamiltonian (eq 29 or eq 36) should also be useful in other mixed quantum-classical methods based on the Meyer-Miller mapping Hamiltonian. We note that the additional computation cost for the commutator variables is negligible in comparison to the force for nuclei for realistic systems. So we expect that the general mapping Hamiltonian with commutator variables (eq 29 or eq 36) will be useful for on-the-fly nonadiabatic dynamics[55, 56]. (See more discussion for mapping dynamics in the adiabatic representation in **Section S1** of Supporting Information.)

The strategy with a commutator variable matrix can in principle be utilized in other mapping models (e.g., those of the unified framework of Ref. [13]). For example, the general mapping Hamiltonian for Model I of Ref. [13] in the diabatic representation is

$$H_{map}\left(\mathbf{R},\mathbf{P};\mathbf{x},\mathbf{p},\mathbf{y},\mathbf{p}_y;\mathbf{\Gamma}\right) = \frac{1}{2}\mathbf{P}^T\mathbf{M}^{-1}\mathbf{P} + \sum_{n,m=1}^{F}\left[(x^{(n)}p_y^{(m)} - y^{(n)}p_x^{(m)}) - \Gamma_{nm}\right]V_{mn}(\mathbf{R}) \quad (54)$$

or

$$\begin{aligned}H_{map}\left(\mathbf{R},\mathbf{P};\mathbf{x},\mathbf{p},\mathbf{y},\mathbf{p}_y;\tilde{\mathbf{x}},\tilde{\mathbf{p}}\right) &= \frac{1}{2}\mathbf{P}^T\mathbf{M}^{-1}\mathbf{P} \\ &+ \sum_{n,m=1}^{F}V_{mn}(\mathbf{R})\left[(x^{(n)}p_y^{(m)} - y^{(n)}p_x^{(m)}) - \sum_{k=1}^{F}\frac{s_k}{2}\left(\tilde{x}_k^{(n)}\tilde{x}_k^{(m)} + \tilde{p}_k^{(n)}\tilde{p}_k^{(m)}\right)\right]\end{aligned}, \quad (55)$$

where $\{x^{(n)}, y^{(n)}; p_x^{(n)}, p_y^{(n)}\}$ are the mapping variables for the $n$-th electronic DOF. The Hamiltonian with commutator variables of eq 55 should lead to more accurate trajectory-based nonadiabatic dynamics than the mapping Hamiltonian of Model I of Ref. [13] in the eCMM approach in Ref. [41] or the SQC approach in Ref. [145]. The isomorphism proposed in Ref. [146] indicates that either eq 54 or eq 29 is a more general phase space mapping Hamiltonian beyond the conventional Li-Miller Hamiltonian for the second-quantized many-electron Hamiltonian[147-150]. Further investigations along this line will shed light on more comprehensive insight for developing phase space mapping approaches for nonadiabatic dynamic processes from photochemistry to electron transfer, as well as for nonequilibrium electronic transport processes, in realistic experimentally related chemical, biological, and materials systems[1-10, 59, 126-136, 151-158].

# ■ ASSOCIATED CONTENT

**Supporting Information**.

Supporting Information is available free of charge via the Internet at the ACS website.

Supporting Information includes four sections: Evolution of nuclear and electronic variables; Results for more values for parameter $\gamma$ in the exact mapping formulation; More results on the FMO monomer; General mapping Hamiltonian with commutator variables in the symmetrical quasi-classical dynamics with triangle window functions; More simulation details of the eCMM/eCMMcv applications. (PDF)

# ■ AUTHOR INFORMATION

**Corresponding Author**

*E-mail: jianliupku@pku.edu.cn

**ORCID**

Xin He: 0000-0002-5189-7204

Baihua Wu: 0000-0002-1256-6859

Zhihao Gong: 0000-0002-9643-633X

Jian Liu: 0000-0002-2906-5858

**Notes**

The authors declare no competing financial interest.

■ **ACKNOWLEDGMENT**

This work was supported by the National Natural Science Foundation of China (NSFC) Grants No. 21961142017, and by the Ministry of Science and Technology of China (MOST) Grant No. 2017YFA0204901. We acknowledge the High-performance Computing Platform of Peking University, Beijing PARATERA Tech CO., Ltd., and Guangzhou supercomputer center for providing computational resources. We thank William H. Miller, Jianshu Cao and Oleg V. Prezhdo for their comments on the manuscript.

11. Meyer, H.-D.; Miller, W. H., A Classical Analog for Electronic Degrees of Freedom in Nonadiabatic Collision Processes. *J. Chem. Phys.* **1979**, *70*, 3214-3223. http://dx.doi.org/10.1063/1.437910

12. Stock, G.; Thoss, M., Semiclassical Description of Nonadiabatic Quantum Dynamics. *Phys. Rev. Lett.* **1997**, *78*, 578-581. http://dx.doi.org/10.1103/PhysRevLett.78.578

13. Liu, J., A Unified Theoretical Framework for Mapping Models for the Multi-State Hamiltonian. *J. Chem. Phys.* **2016**, *145*, 204105. http://dx.doi.org/10.1063/1.4967815

14. Sun, X.; Wang, H. B.; Miller, W. H., Semiclassical Theory of Electronically Nonadiabatic Dynamics: Results of a Linearized Approximation to the Initial Value Representation. *J. Chem. Phys.* **1998**, *109*, 7064-7074. http://dx.doi.org/10.1063/1.477389

15. Coronado, E. A.; Batista, V. S.; Miller, W. H., Nonadiabatic Photodissociation Dynamics of Icn in the (a)over-Tilde Continuum: A Semiclassical Initial Value Representation Study. *J. Chem. Phys.* **2000**, *112*, 5566-5575. http://dx.doi.org/10.1063/1.481130

16. Ananth, N.; Venkataraman, C.; Miller, W. H., Semiclassical Description of Electronically Nonadiabatic Dynamics Via the Initial Value Representation. *J. Chem. Phys.* **2007**, *127*, 084114. http://dx.doi.org/10.1063/1.2759932

17. Tao, G.; Miller, W. H., Semiclassical Description of Electronic Excitation Population Transfer in a Model Photosynthetic System. *J. Phys. Chem. Lett.* **2010**, *1*, 891-894. http://dx.doi.org/10.1021/jz1000825
42

30. Muller, U.; Stock, G., Flow of Zero-Point Energy and Exploration of Phase Space in Classical Simulations of Quantum Relaxation Dynamics. II. Application to Nonadiabatic Processes. *J. Chem. Phys.* **1999**, *111*, 77-88. http://dx.doi.org/10.1063/1.479255

31. Thoss, M.; Stock, G., Mapping Approach to the Semiclassical Description of Nonadiabatic Quantum Dynamics. *Phys. Rev. A* **1999**, *59*, 64-79. http://dx.doi.org/10.1103/PhysRevA.59.64

32. Stock, G.; Thoss, M., Classical Description of Nonadiabatic Quantum Dynamics. In *Advances in Chemical Physics, Vol 131*, Rice, S. A., Ed. 2005; Vol. 131, pp 243-375.

33. Golosov, A. A.; Reichman, D. R., Classical Mapping Approaches for Nonadiabatic Dynamics: Short Time Analysis. *J. Chem. Phys.* **2001**, *114*, 1065-1074. http://dx.doi.org/10.1063/1.1332812

34. Bonella, S.; Coker, D. F., Land-Map, a Linearized Approach to Nonadiabatic Dynamics Using the Mapping Formalism. *J. Chem. Phys.* **2005**, *122*, 194102. http://dx.doi.org/10.1063/1.1896948

35. Saller, M. A. C.; Kelly, A.; Richardson, J. O., On the Identity of the Identity Operator in Nonadiabatic Linearized Semiclassical Dynamics. *J. Chem. Phys.* **2019**, *150*, 071101. http://dx.doi.org/10.1063/1.5082596

36. Ananth, N.; Miller, T. F., III, Exact Quantum Statistics for Electronically Nonadiabatic Systems Using Continuous Path Variables. *J. Chem. Phys.* **2010**, *133*, 234103. http://dx.doi.org/10.1063/1.3511700
45

**TOC Graph**

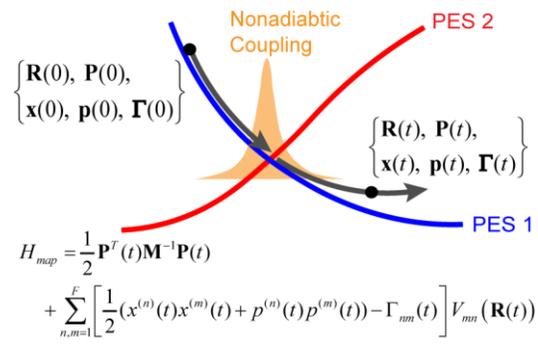

# Supporting Information: Commutator Matrix in Phase Space Mapping Models for Nonadiabatic Quantum Dynamics

*Xin He, Baihua Wu, Zhihao Gong, and Jian Liu\**

Beijing National Laboratory for Molecular Sciences, Institute of Theoretical and Computational Chemistry, College of Chemistry and Molecular Engineering, Peking University, Beijing 100871, China

AUTHOR INFORMATION

**Corresponding Author**

* Electronic mail: jianliupku@pku.edu.cn

## S1. Evolution of nuclear and electronic variables

### S1-A Evolution of electronic mapping variables and auxiliary commutator variable matrix

Rather than employing eq 38 of the main text of the paper for the evolution of the electronic mapping DOFs $\{\mathbf{x},\mathbf{p}\}$ and auxiliary variables $\{\tilde{\mathbf{x}},\tilde{\mathbf{p}}\}$ for the commutator matrix, a more compact and convenient approach is using the complex vector representation of the electronic DOFs, i.e.,

$$\mathbf{g}(t) = (\mathbf{x}(t) + i\mathbf{p}(t))/\sqrt{2} \quad , \tag{S1}$$

as well as the complex matrix representation of the auxiliary variables, i.e., $\mathbf{\Gamma}(t)$.

We assume that the $F$ electronic states form a complete basis set, which has already been employed in the main text. In the general representation, it states that the identity operator in the electronic state space is

$$\hat{I}_{ele}(\mathbf{R}) = \sum_{n=1}^{F} |n(\mathbf{R})\rangle\langle n(\mathbf{R})| \tag{S2}$$

It is not difficult to show that an equivalent form of the Hamiltonian in quantum mechanics is

$$\hat{H} = \frac{1}{2}\hat{\mathbf{P}}^T \mathbf{M}^{-1}\hat{\mathbf{P}} + \sum_{n,m=1}^{F} V_{mn}^{\mathit{eff}}\left(\hat{\mathbf{R}},\hat{\mathbf{P}}\right)|n\rangle\langle m| \tag{S3}$$

in the general representation. Here, the element in row $k$ and column $n$ of the $F \times F$ matrix, $\mathbf{V}^{\mathit{eff}}$, is

$$V_{kn}^{\mathit{eff}}(\mathbf{R},\mathbf{P}) = -\left(i\hbar \sum_{j=1}^{N} d_{kn}^{(j)}(\mathbf{R})\frac{P_j}{M_j} + \hbar^2 \sum_{j=1}^{N} \frac{1}{2M_j}\overline{\overline{d}}_{kn}^{(j)}(\mathbf{R})\right) + V_{kn}(\mathbf{R}) \tag{S4}$$

for the general representation. In eq (S4), $d_{kn}^{(j)}(\mathbf{R}) = \left\langle \phi_k(\mathbf{R}) \left| \frac{\partial}{\partial R_j} \right| \phi_n(\mathbf{R}) \right\rangle$ is the first-order nonadiabatic coupling vector, $\overline{\overline{d}}_{kn}^{(j)}(\mathbf{R}) = \left\langle \phi_k(\mathbf{R}) \left| \frac{\partial^2}{\partial R_j^2} \right| \phi_n(\mathbf{R}) \right\rangle$ is the second-order coupling term between $k$-th and $n$-th electronic states, and $V_{kn}(\mathbf{R})$ is the element in row $k$ and column $n$ of the multi-state potential matrix (with $R_j$ the coordinate of the $j$-th nuclear DOF). The total number of nuclear DOFs is $N$ as defined in the main text.

Specifically, eq (S4) becomes

$$V_{kn}^{\text{eff}} = V_{kn}(\mathbf{R}) \tag{S5}$$

in the diabatic representation, and

$$V_{kn}^{\text{eff}} = E_n(\mathbf{R})\delta_{kn} - \left( i\hbar \sum_{j=1}^{N} d_{kn}^{(j)}(\mathbf{R}) \frac{P_j}{M_j} + \hbar^2 \sum_{j=1}^{N} \frac{1}{2M_j} \overline{\overline{d}}_{kn}^{(j)}(\mathbf{R}) \right) \tag{S6}$$

in the adiabatic representation, where $E_n(\mathbf{R}) = V_{nn}(\mathbf{R})$ is the adiabatic potential surface. (More discussions on why the direct evaluation of $\overline{\overline{d}}_{kn}^{(j)}(\mathbf{R})$ is *not* necessary are provided in **Section S1-C.**)

The phase space mapping model for the Hamiltonian eq (S3) in the general representation implies the same constraint of eq 6 or eq 28 of the main text. The exact phase space mapping formulation of Section 2.1 of the main text applies to the general representation. Define the time evolution matrix in each time step

$$\mathbf{U}(\mathbf{R},\mathbf{P};\Delta t) = \exp\left[-i\mathbf{V}^{\text{eff}}(\mathbf{R},\mathbf{P})\Delta t / \hbar \right] . \tag{S7}$$

It is straightforward to obtain an equivalent formulation for the integrator of the equations of motion eq 38 in the main text of the paper,

$$\mathbf{g}(t+\Delta t) = \mathbf{U}(\mathbf{R},\mathbf{P};\Delta t)\mathbf{g}(t) \tag{S8}$$

$$\mathbf{\Gamma}(t+\Delta t) = \mathbf{U}(\mathbf{R},\mathbf{P};\Delta t)\mathbf{\Gamma}(t)\mathbf{U}^\dagger(\mathbf{R},\mathbf{P};\Delta t) \ . \tag{S9}$$

Here the initial condition for the real and imaginary parts of $\mathbf{g}(0)$ are uniformly sampled on constraint space $\mathcal{S}(\mathbf{x},\mathbf{p})$ (see details in the main text). The initial condition $\mathbf{\Gamma}(0)$ for the commutator matrix, which is determined by eq 44 of the main text of the paper, is

$$\Gamma_{nm}(0) = (|g_n(0)|^2 - \delta_{n,j_{occ}})\delta_{nm} \ . \tag{S10}$$

where $g_n(0) = (x^{(n)}(0) + ip^{(n)}(0))/\sqrt{2}$ is the $n$-th element of vector $\mathbf{g}(0)$.

### S1-B Evolution of nuclear variables

We adopt the same strategy as proposed by Cotton and Miller in Ref. [1] to extend the phase space mapping approaches in the adiabatic representation, where the mapping Hamiltonian in the general representation reads,

$$\begin{aligned}H_{map}(\mathbf{R},\mathbf{P};\mathbf{x},\mathbf{p};\mathbf{\Gamma}) = &\frac{1}{2}\left(\mathbf{P} - \sum_{nm}\left(x^{(m)}p^{(n)} + \Gamma_{mn}^i\right)\mathbf{d}_{nm}\right)^T \mathbf{M}^{-1}\left(\mathbf{P} - \sum_{nm}\left(x^{(m)}p^{(n)} + \Gamma_{mn}^i\right)\mathbf{d}_{nm}\right) \\ &+ \sum_{nm}\left[\frac{1}{2}\left(x^{(m)}x^{(n)} + p^{(m)}p^{(n)}\right) - \Gamma_{mn}^r\right]V_{nm}(\mathbf{R})\end{aligned} \tag{S11}$$

In eq (S11), $\mathbf{d}_{nm} = 0$ in the diabatic representation, while $V_{nm}(\mathbf{R}) = E_n(\mathbf{R})\delta_{nm}$ in the adiabatic representation. $\Gamma_{mn}^i = \sum_{k=1}^{F}\frac{s_k}{2}\left(\tilde{p}_k^{(m)}\tilde{x}_k^{(n)} - \tilde{x}_k^{(m)}\tilde{p}_k^{(n)}\right)$ and $\Gamma_{mn}^r = \sum_{k=1}^{F}\frac{s_k}{2}\left(\tilde{x}_k^{(m)}\tilde{x}_k^{(n)} + \tilde{p}_k^{(m)}\tilde{p}_k^{(n)}\right)$ are the imaginary and real parts of $\Gamma_{mn}$, i.e., the element in row $m$ and column $n$ of the commutator matrix. In the same spirit of the work of Cotton and Miller[1], kinematic momentum $\mathbf{P}_{kin} \triangleq \mathbf{M}\dot{\mathbf{R}} = \mathbf{P} - \sum_{nm}\left(x^{(m)}p^{(n)} + \Gamma^{i}_{mn}\right)\mathbf{d}_{nm}$ can be introduced to yield a new form of $\mathbf{V}^{\text{eff}}$ equivalent to eq (S4),

$$V^{\text{eff}}_{kn}(\mathbf{R}, \mathbf{P}_{kin}) = -i\hbar \sum_{j=1}^{N} d^{(j)}_{kn}(\mathbf{R}) \frac{P_{kin,j}}{M_j} + V_{kn}(\mathbf{R}) \ . \tag{S12}$$

In the adiabatic representation, the equations of motion for the nuclear DOFs are

$$\begin{aligned}\dot{\mathbf{R}} &= \mathbf{M}^{-1}\mathbf{P}_{kin} \\ \dot{\mathbf{P}}_{kin} &= -\sum_{n,m=1}^{F}\left[\frac{1}{2}\left(x^{(m)}x^{(n)} + p^{(m)}p^{(n)}\right) - \Gamma^{r}_{mn}\right]\left[\partial_{\mathbf{R}}E_n(\mathbf{R})\delta_{nm} + (E_m(\mathbf{R}) - E_n(\mathbf{R}))\mathbf{d}_{nm}(\mathbf{R})\right]\end{aligned} \ . \tag{S13}$$

We note that the value of kinetic momentum $\mathbf{P}_{kin}$ in the adiabatic representation is equivalent to that of the canonical momentum in the corresponding diabatic representation. An effective force matrix can be introduced, i.e.,

$$\mathbf{F}_{nm}(\mathbf{R}) = \partial_{\mathbf{R}}V_{nm}(\mathbf{R}) + \sum_{k}\left(\mathbf{d}_{nk}(\mathbf{R})V_{km}(\mathbf{R}) - V_{nk}(\mathbf{R})\mathbf{d}_{km}(\mathbf{R})\right) \ . \tag{S14}$$

It is easy to show $\mathbf{F}_{nm}(\mathbf{R}) = \partial_{\mathbf{R}}V_{nm}(\mathbf{R})$ in the diabatic representation while $\mathbf{F}_{nm}(\mathbf{R}) = \partial_{\mathbf{R}}E_n(\mathbf{R})\delta_{nm} + \left(E_m(\mathbf{R}) - E_n(\mathbf{R})\right)\mathbf{d}_{nm}(\mathbf{R})$ in the adiabatic representation. With the effective force matrix, mapping variables, and commutator variables, the form for eq (S13) in the general representation reads

$$\begin{aligned}\dot{\mathbf{R}} &= \mathbf{M}^{-1}\mathbf{P}_{kin} \\ \dot{\mathbf{P}}_{kin} &= -\sum_{n,m=1}^{F}\frac{1}{2}\left(x^{(m)}x^{(n)} + p^{(m)}p^{(n)} - \Gamma^{r}_{mn}\right)\mathbf{F}_{nm}(\mathbf{R})\end{aligned} \ . \tag{S15}$$

The integrator for each time step $\Delta t$ can be constructed from eq (S8), eq (S9) and eq (S15). For example, an integrator scheme that we use in the paper reads

$$\mathbf{P}_{kin}(t+\Delta t/2) = \mathbf{P}_{kin}(t) - \sum_{n,m=1}^{F}\left[\frac{1}{2}\left(x^{(m)}(t)x^{(n)}(t) + p^{(m)}(t)p^{(n)}(t)\right)\right.$$
$$\left. - \Gamma_{mn}^{r}(t)\right]\mathbf{F}_{nm}(\mathbf{R}(t))\Delta t/2$$
$$\mathbf{R}(t+\Delta t) = \mathbf{R}(t) + \mathbf{M}^{-1}\mathbf{P}_{kin}(t+\Delta t/2)$$
$$\begin{cases}\mathbf{g}(t+\Delta t) = \mathbf{U}(\mathbf{R}(t+\Delta t), \mathbf{P}_{kin}(t+\Delta t/2); \Delta t)\mathbf{g}(t) \\ \mathbf{\Gamma}(t+\Delta t) = \mathbf{U}(\mathbf{R}(t+\Delta t), \mathbf{P}_{kin}(t+\Delta t/2); \Delta t)\mathbf{\Gamma}(t)\mathbf{U}^{\dagger}(\mathbf{R}(t+\Delta t), \mathbf{P}_{kin}(t+\Delta t/2); \Delta t)\end{cases} \quad . \quad (S16)$$
$$\mathbf{P}_{kin}(t+\Delta t) = \mathbf{P}_{kin}(t+\Delta t/2) - \sum_{n,m=1}^{F}\left[\frac{1}{2}\left(x^{(m)}(t+\Delta t)x^{(n)}(t+\Delta t) + p^{(m)}(t+\Delta t)p^{(n)}(t+\Delta t)\right)\right.$$
$$\left. - \Gamma_{mn}^{r}(t+\Delta t)\right]\mathbf{F}_{nm}(\mathbf{R}(t+\Delta t))\Delta t/2$$

**S1-C More Discussions on the nonadiabatic coupling**

Consider matrix $\mathbf{d}^{(j)}(\hat{\mathbf{R}})$, of which the element, $d_{kn}^{(j)}(\mathbf{R})$, is already used in eq (S4). It is easy to prove the commutation relation

$$[\hat{P}_j, \mathbf{d}^{(j)}(\hat{\mathbf{R}})] = -i\hbar\frac{\partial}{\partial R_j}\mathbf{d}^{(j)}(\hat{\mathbf{R}}) \tag{S17}$$

as well as

$$\overline{\overline{d}}_{kn}^{(j)}(\hat{\mathbf{R}}) = \left\langle\phi_k(\hat{\mathbf{R}})\left|\frac{\partial^2}{\partial R_j^2}\right|\phi_n(\hat{\mathbf{R}})\right\rangle = \sum_{m=1}^{F}d_{km}^{(j)}(\hat{\mathbf{R}})d_{mn}^{(j)}(\hat{\mathbf{R}}) + \frac{\partial}{\partial R_j}d_{kn}^{(j)}(\hat{\mathbf{R}}) \quad . \tag{S18}$$

We can then show the equality

$$\left(\hat{P}_j - i\hbar\mathbf{d}^{(j)}(\hat{\mathbf{R}})\right)^2 = \hat{P}_j^2 - 2i\hbar\mathbf{d}^{(j)}(\hat{\mathbf{R}})\hat{P}_j - \hbar^2\overline{\overline{\mathbf{d}}}^{(j)}(\hat{\mathbf{R}}) \quad . \tag{S19}$$

The Wigner-Weyl correspondence suggests

$$i\hat{P}_j\mathbf{d}^{(j)}(\hat{\mathbf{R}}) + i\mathbf{d}^{(j)}(\hat{\mathbf{R}})\hat{P}_j \mapsto 2iP_j\mathbf{d}^{(j)}(\mathbf{R})$$
$$[\hat{P}_j, \mathbf{d}^{(j)}(\hat{\mathbf{R}})] \mapsto 0 \tag{S20}$$

The Wigner function, i.e., the mapping function in Wigner phase space for the nuclear DOFs, for operator $\left(\hat{P}_j - i\hbar\mathbf{d}^{(j)}(\hat{\mathbf{R}})\right)^2$, is

$$P_j^2 - 2i\hbar\mathbf{d}^{(j)}(\mathbf{R})P_j - \hbar^2 \mathbf{d}^{(j)}(\mathbf{R})\mathbf{d}^{(j)}(\mathbf{R}), \tag{S21}$$

instead of the naïve classical correspondence

$$P_j^2 - 2i\hbar\mathbf{d}^{(j)}(\mathbf{R})P_j - \hbar^2 \overline{\overline{\mathbf{d}}}^{(j)}(\mathbf{R}) \quad . \tag{S22}$$

Employing Wigner phase space for the nuclear DOFs avoids the use of the second-order nonadiabatic coupling term. For instance, the expression of eq (S4) in Wigner phase space is simply

$$V_{kn}^{\text{eff}}(\mathbf{R},\mathbf{P}) = -\left(i\hbar\sum_{j=1}^{N} d_{kn}^{(j)}(\mathbf{R})\frac{P_j}{M_j} + \hbar^2\sum_{j=1}^{N}\frac{1}{2M_j}\sum_{m=1}^{F} d_{km}^{(j)}(\mathbf{R})d_{mn}^{(j)}(\mathbf{R})\right) + V_{kn}(\mathbf{R}) \quad . \tag{S23}$$

Provided that phase variables $(\mathbf{R},\mathbf{P})$ are fixed in eq (S23), i.e., in the frozen nuclei limit, the exact evolution of the electronic DOFs does NOT request the second-order nonadiabatic coupling term (even in the adiabatic representation). This demonstrates the advantage of the utilization of Wigner phase space for mapping the nuclear DOFs, especially when (on the fly) *ab initio* calculations are involved.

When we study the equations of motion of the nuclear phase space variables, we can extend the strategy of using the kinematic momentum by Cotton and Miller in Ref. [1] to the eCMM/eCMMcv approaches. As discussed in **Section S1-B**, the second-order nonadiabatic coupling term is avoided too, even when both nuclear dynamics and electronic dynamics are considered.

## S2. Results for more values for parameter $\gamma$ in the exact mapping formulation

We test eCMM, as well as eCMMcv, with more values for parameter $\gamma$ in the exact mapping kernel of the correlation function. A gas phase model [Tully's simple avoided crossing (SAC) model of Section 3.1] and a condensed phase model [the two-level atom-in-cavity model of Section 3.4] are used for demonstration.

### S2-A  More results for Tully's simple avoided crossing model

We use Tully's SAC model to test the performance of eCMM or eCMMcv with more values for parameter $\gamma$ in the exact mapping formulation of the correlation function. Three more values $\{-0.2, 0, 1\}$ are selected for parameter $\gamma$. We use the diabatic representation for demonstration. The results for the transmission coefficients of Tully's SAC model are shown in Figure S1. Panels S1(a) and S1(b) suggest that the eCMM results are relatively sensitive to the value of parameter $\gamma$ in the region where the initial average momentum $P_0$ is below 10 a.u. In comparison, Panels S1(c) and S1(d) indicate that the eCMMcv results are much more robust as for the value of parameter $\gamma$ varies. This is consistent with the comparison between eCMM and eCMMcv in Figures 1-2 (for the same SAC model) in the paper. We then conclude that eCMMcv performs better than eCMM for this standard gas phase model.

### S2-B  More results for the two-level atom-in-cavity model

We use the two-level atom-in-cavity model as an example of the condensed phase system to test eCMM as well as eCMMcv with more values for parameter $\gamma$ in the exact mapping kernel of the correlation function. Two more values, $\gamma=0$ and $\gamma=1$, are used. Results are presented in Figure

S2. In comparison of Figure S2 to Figure 12 for the same model, it implies that the performance of eCMM or eCMMcv is relatively insensitive to the parameter for $\left(\sqrt{F+1}-1\right)/F \leq \gamma \leq 1/2$. We then suggest that $\left[\left(\sqrt{F+1}-1\right)/F, 1/2\right]$ should be the region for parameter $\gamma$ for all types of systems.

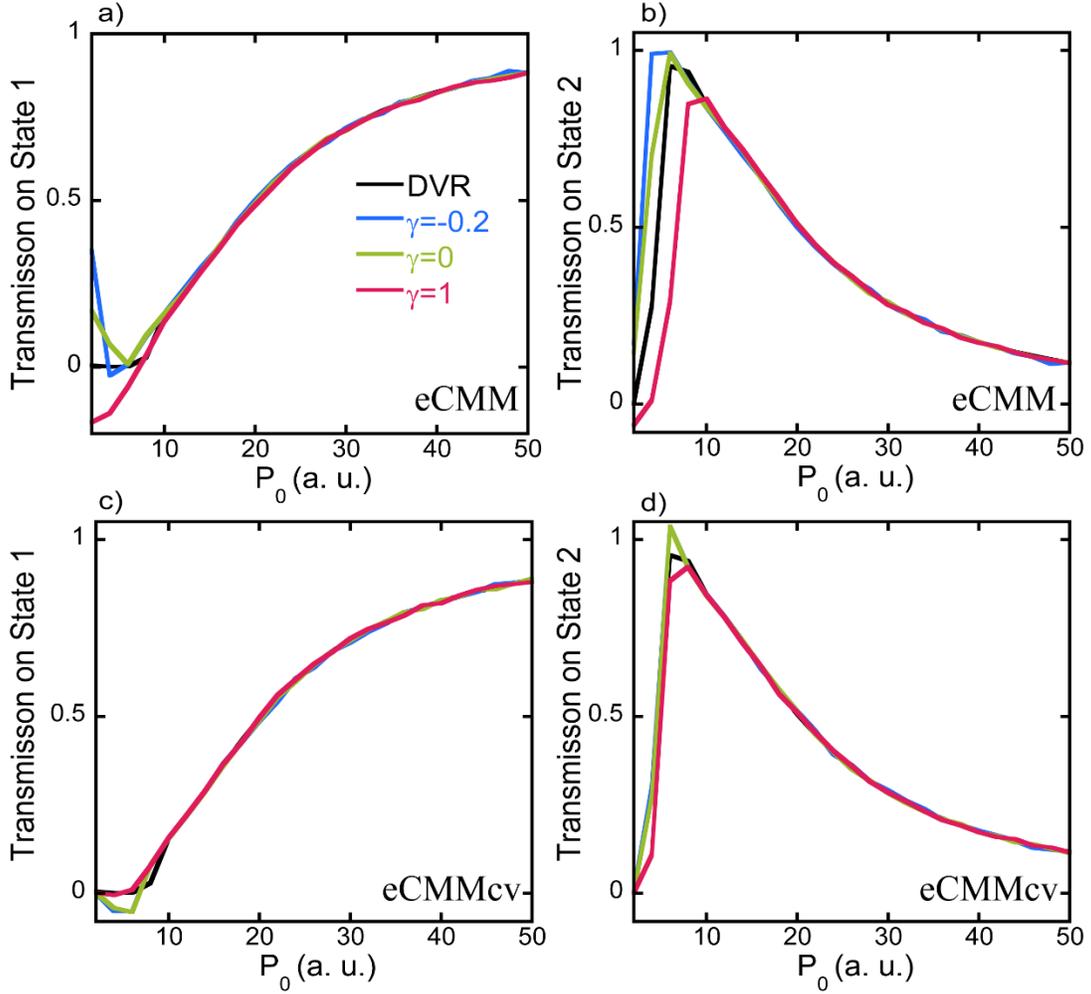

**Figure S1**: Results for Tully's SAC model with parameter $\gamma$ in $\{-0.2, 0, 1\}$. Panels (a) and (b) show the transmission coefficients produced by eCMM on state 1 and state 2, respectively. Panels (c) and (d) demonstrate those yielded by eCMMcv. Solid black line denotes DVR, and Blue, and Green, Red solid lines stand for results for parameter $\gamma$ in $\{-0.2, 0, 1\}$, respectively.

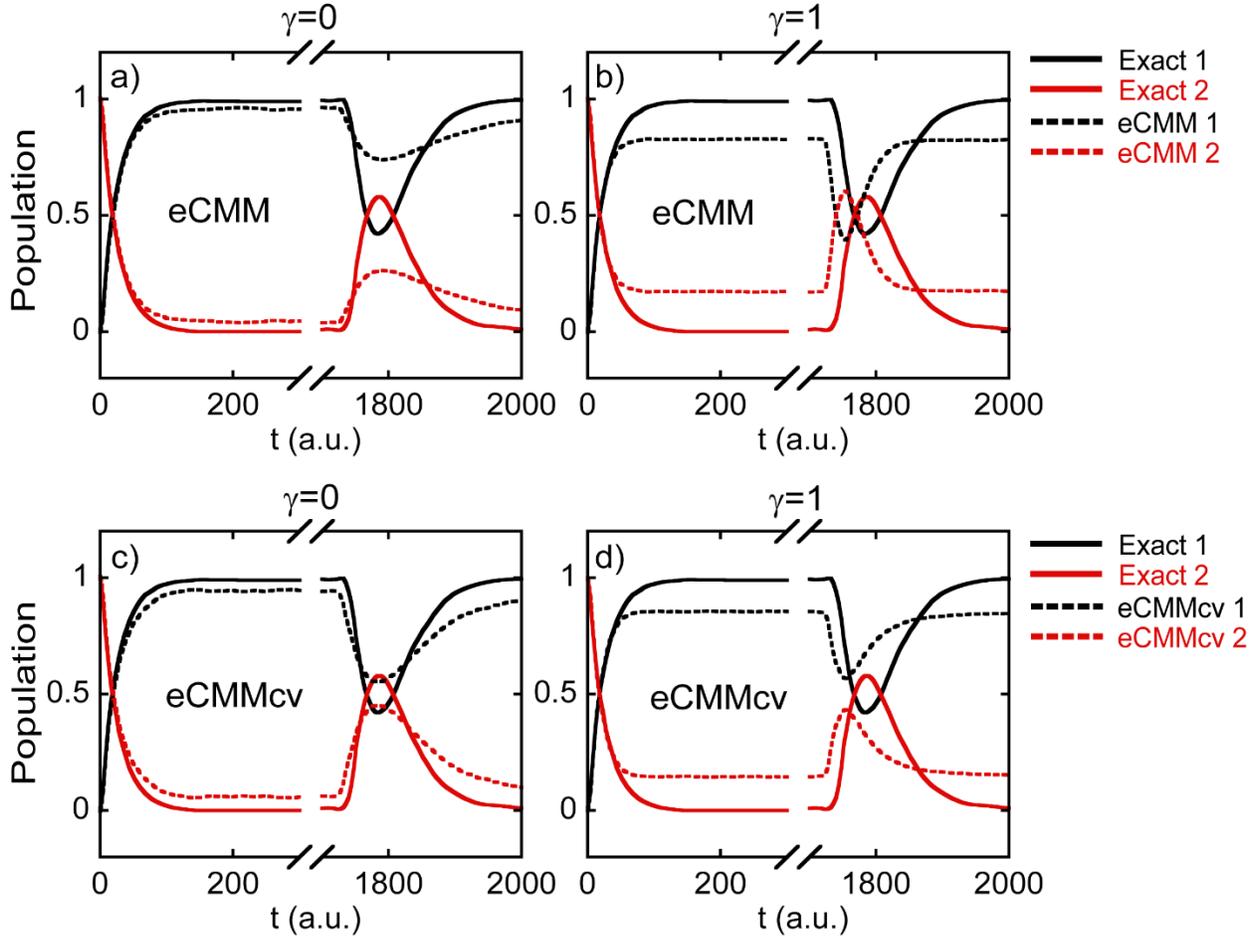

**Figure S2**: Results for the two-level atomic system in optical cavity. The Black (Red) color stands for the population on state 1 (state 2). Panels S2(a) and S2(b) show the eCMM results with parameter $\gamma$ in $\{0, 1\}$, respectively, while Panels S2(c) and S2(d) do so for the eCMMcv results. Solid lines: exact benchmark data from Refs. [2,3]. Dashed lines: eCMM or eCMMcv results.

## S3   More results for the FMO monomer

Here we present more results for the FMO monomer model of Section 3.2 of the main text. We first compare eCMM/eCMMcv to HEOM on the long time dynamic behavior at 77K, and at 300K, respectively, when pigment 1/site 1 of the FMO monomer is initially activated. Figure S3 demonstrates the results at 77K. In comparison to the exact results obtained by HEOM, the eCMM/eCMMcv approaches perform overall reasonably well for parameter $\gamma \in \left[ \left( \sqrt{F+1} - 1 \right)/F, 1/2 \right]$. For instance, $\gamma = \left( \sqrt{F+1} - 1 \right)/F \approx 0.261$ works better for the population dynamics of site 3, while $\gamma = 0.5$ does better for that of site 1. In contrast, as shown in Panel S3(e), Ehrenfest dynamics performs poorly for the long time limit. In the right subpanel of Panel S3(e) for the long time limit results, Ehrenfest dynamics tends to yield nearly close results for all sites/pigments. This is intrinsically unphysical, in comparison to the benchmark results produced by HEOM.

Figure S4 presents the results for the same case, but at the room temperature. As the temperature increases to 300K, eCMM/eCMMcv perform even better, predicting results that are much closer to benchmark data. This indicates that eCMM/eCMMcv are useful in studying exciton dynamics or nonadiabatic dynamics in biological systems under ambient conditions. In contrast, Ehrenfest dynamics lead to significantly poor results in the long time limit at the room temperature—the sites (pigments) are nearly equally populated! Such a long time limit behavior of Ehrenfest dynamics is substantially different from that suggested by eCMM/eCMMcv or HEOM.

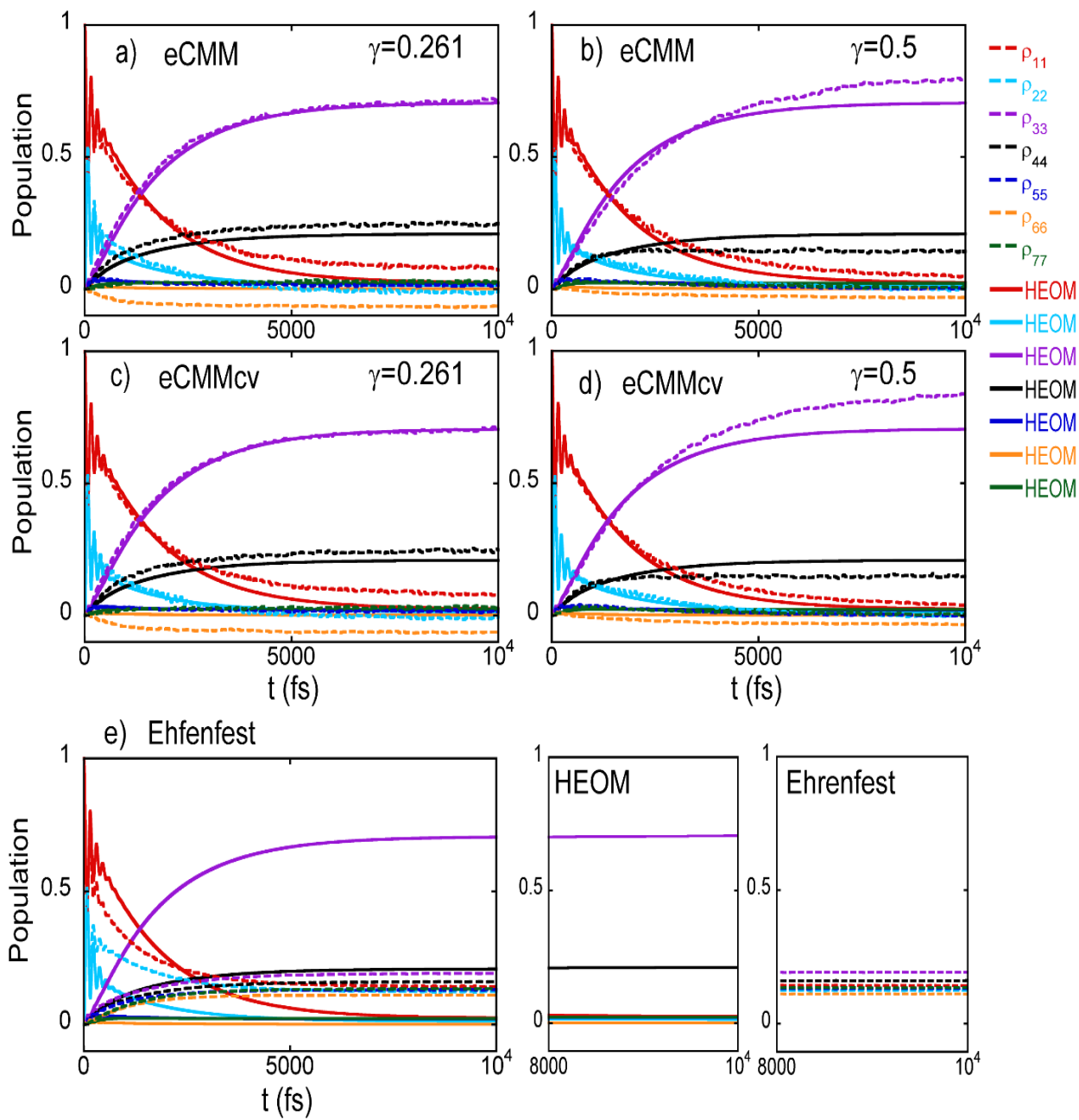

**Figure S3**: Results for longer time population dynamics for the FMO monomer model at 77K when pigment 1/site 1 is initially activated. All markers are the same as **Figure 7** of the main text. Red, Blue, Purple, Black, Blue, Orange and Green lines present populations of site 1,2,3,4,5,6, and 7, respectively. Dashed lines: The eCMM results in Panels (a)-(b), or the eCMMcv results in Panels (c)-(d); Ehrenfest results for Panel (e). Solid lines: Exact results obtained by HEOM.

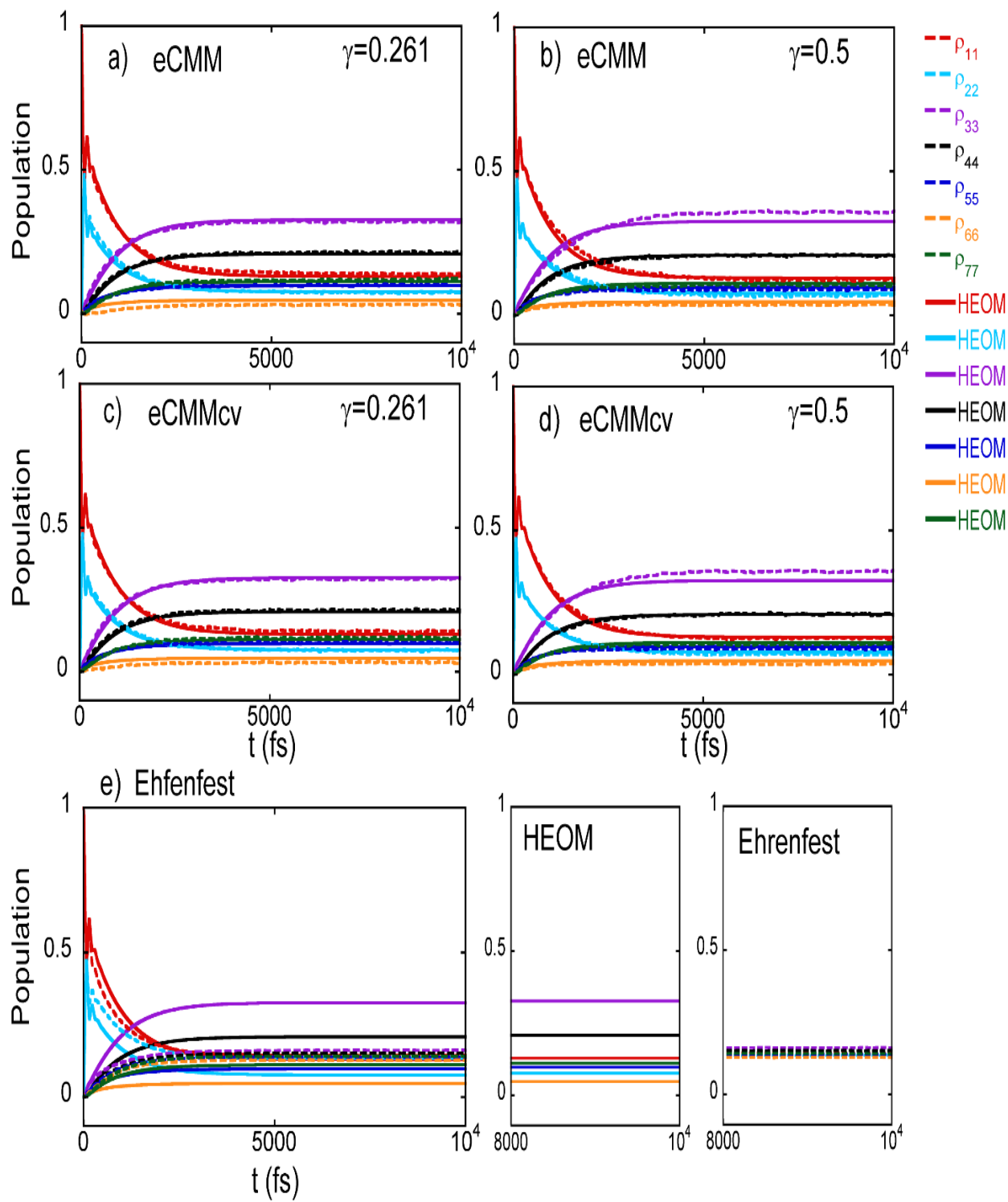

**Figure S4**: Long time limit for the population dynamics of the FMO monomer at 300K, when pigment 1/site 1 is initially activated. All markers are same as **Figure S3**.

## S4. General mapping Hamiltonian with commutator variables in the symmetrical quasi-classical dynamics with triangle window functions

The general mapping Hamiltonian with commutator variables [eq 36 of the main text] can be used to replace the original Meyer-Miller mapping Hamiltonian [eq 3 of the main text] in the symmetrical quasi-classical (SQC) dynamics with triangle window functions of Cotton and Miller. We utilize the latest version of SQC of Cotton and Miller of Ref. [4] to test the performance of the general mapping Hamiltonian with commutator variables.

We compare the original Meyer-Miller mapping Hamiltonian [eq 3 of the main text] to the general mapping Hamiltonian with commutator variables [eq 36 of the main text] in the latest SQC method of Ref. [4]. The latter is denoted SQCcv (symmetrical quasi-classical approach with commutator variables) for convenience. Figure S5 presents the results for two spin-boson models with the Debye bath, where the spectral density reads

$$J(\omega) = 2\lambda \frac{\omega_c \omega}{\omega_c^2 + \omega^2} \quad . \tag{S24}$$

Here $\omega_c$ is the characteristic frequency of the bath, and $\lambda$ denotes the reorganization energy. The spin-boson Hamiltonian operator reads

$$\hat{H} = \varepsilon \hat{\sigma}_z + \Delta \hat{\sigma}_x + \left( \sum_i c_i \hat{R}_i \right) \hat{\sigma}_z + \sum_i \frac{1}{2} \left( \hat{P}_i^2 + \omega_i^2 \hat{R}_i^2 \right) \quad , \tag{S25}$$

where $\hat{\sigma}_x$ and $\hat{\sigma}_z$ are Pauli matrices in the $x$ and $z$ directions, respectively, $\varepsilon$ represents the detuning between states $|1\rangle$ and $|2\rangle$, $\Delta$ denotes the tunneling amplitude, and $\{\hat{R}_i, \hat{P}_i\}$ are the mass-weighted position and momentum operators of the $i$-th bath oscillator ($1 \leq i \leq N_b$), respectively. Frequencies and coupling strengths $\{\omega_i, c_i\}$ are sampled from the discretization procedure for the Debye bath as described in Ref. [5]. $\Delta = 1$ and $\varepsilon = 1$ are used for the spin system. A relative low temperature $\beta = 5$ in the nonadiabatic region ($\omega_c = 5\Delta$) is used for the spin-boson models. The initial occupied state is state 1, with no correlation with the bosonic bath, i.e., the initial density is $|1\rangle\langle 1| \otimes e^{-\beta \hat{H}_b}/Z_b$, where $\hat{H}_b = \sum_i \frac{1}{2}\left(\hat{P}_i^2 + \omega_i^2 \hat{R}_i^2\right)$ is the bare bath Hamiltonian and $Z_b = \text{Tr}_n\left[e^{-\beta \hat{H}_b}\right]$ the partition function for the bath.

As demonstrated in Figure S5(a), SQCcv is able to improve over the original SQC method of of Cotton and Miller of Ref. [4] for the oscillation behavior at short times. Figure S5(b) shows that the SQCcv results are closer to the exact asymptotic long-time limit than the SQC ones in the challenging case.

Figure S6 studies nonadiabatic dynamics through the conical intersection in the pyrazine model with two electronic states and three vibrational modes of Ref. [6]. While SQCcv and SQC lead to similar long time dynamics behaviors that are reasonably good, it is evident that SQCcv yields more accurate short time dynamics than SQC.

Figure S5 and Figure S6 suggest that the performance of the successful SQC approach can further be improved when the general mapping Hamiltonian with commutator variables (eq 36 of the main text) is employed to replace the original Meyer-Miller mapping Hamiltonian. This is consistent with the conclusion the overall performance of eCMMcv is better than that of eCMM. It will be interesting to investigate the behavior of the general mapping Hamiltonian with commutator variables (eq 36 of the main text) in other nonadiabatic approaches based on the Meyer-Miller mapping Hamiltonian model.

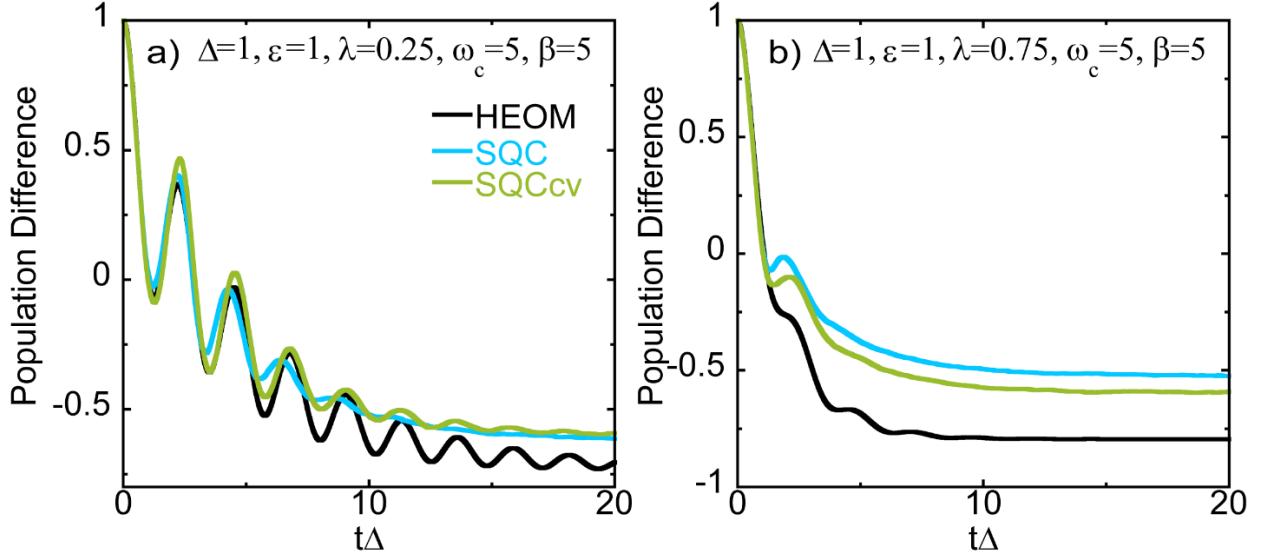

**Figure S5:** Difference between the population of state 1 and that of state 2 of two asymmetrical spin-boson models with the Debye bath. The bath parameters in the two panels are a) $\lambda = 0.25$, $\omega_c = 5$, $\beta = 5$ and b) $\lambda = 0.75$, $\omega_c = 5$, $\beta = 5$, respectively. The number of bath modes is $N_b = 200$. Black, Cyan and Green lines present the results of HEOM, SQC (of Ref. [4]), and SQCcv, respectively. The case in Panel b) is challenging.

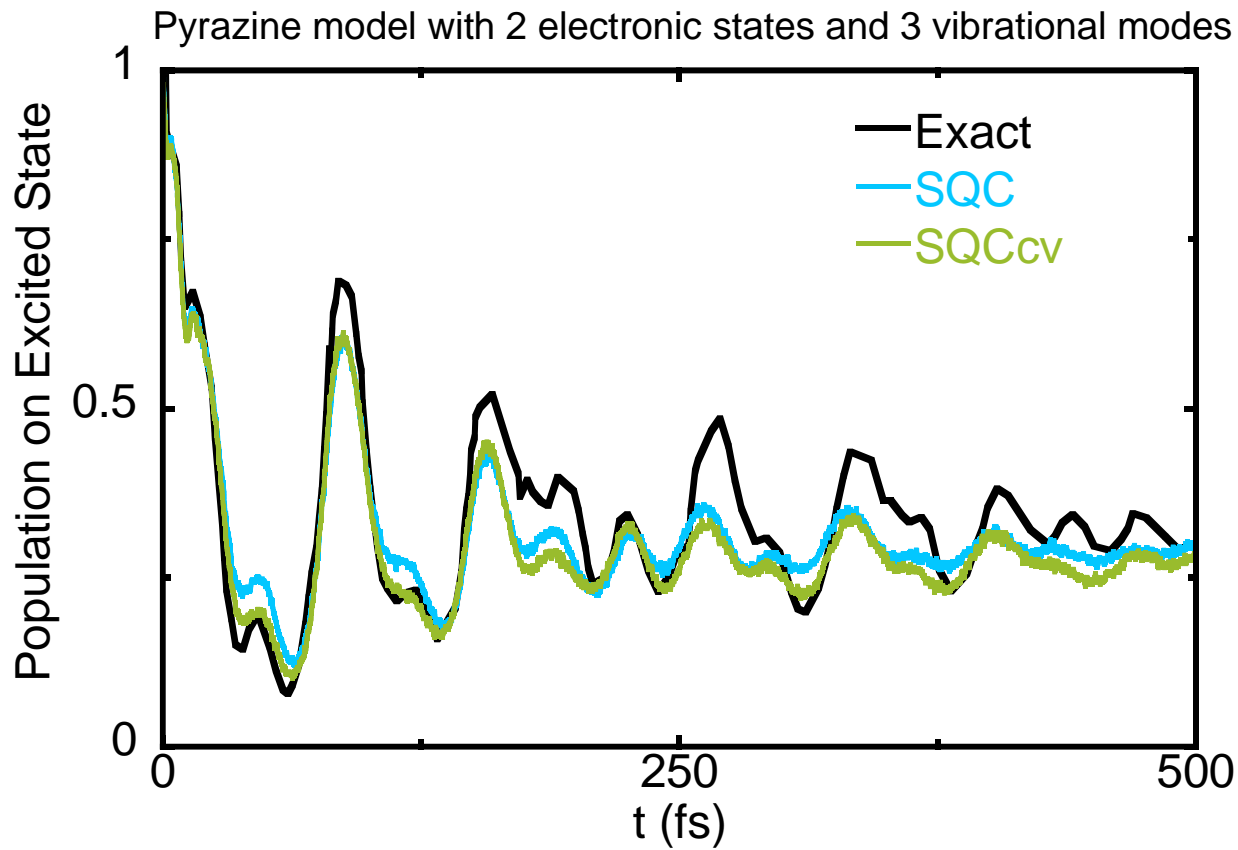

**Figure S6:** Population dynamics of the excited state of the pyrazine model. Parameters of the model are listed in Ref. [6]. Black, Cyan and Green lines stand for the results produced by ML-MCTDH from Ref. [7], SQC (of Ref. [4]), and SQCcv, respectively.

## S5. More simulation details of the eCMM/eCMMcv applications

Table S1 lists the time step size as well as the total number of trajectories that are used to guarantee full convergence, for each type of models studied in the main text.

**Table S1: Parameters for the eCMM/eCMMcv approaches**

| Model | Time Step Size | Number of Trajectories |
|---|---|---|
| SAC/DAC | 0.05(a.u.) | $\sim 10^5$ |
| photo-dissociation models | 0.02(a.u.) | $\sim 10^5$ |
| FMO | 0.1(fs) | $\sim 10^5$ |
| Atom-in-Cavity models | 0.1(a.u.) | $\sim 10^5$ |

■ **ACKNOWLEDGMENT**

This work was supported by the National Natural Science Foundation of China (NSFC) Grants No. 21961142017, and by the Ministry of Science and Technology of China (MOST) Grant No. 2017YFA0204901. We acknowledge the High-performance Computing Platform of Peking University, Beijing PARATERA Tech CO., Ltd., and Guangzhou supercomputer center for providing computational resources.